\documentclass[12pt]{article}
\usepackage{epsfig,graphicx,bm,amsmath}

\newcommand{\abs}[1]{\lvert#1\rvert}

\newcommand{\IP}[1]{\langle#1\rangle}

\newcommand{\be}{\begin{equation}}
\newcommand{\bea}{\begin{eqnarray}}
\newcommand{\eea}{\end{eqnarray}}
\newcommand{\ba}{\begin{array}}
\newcommand{\ea}{\end{array}}
\newcommand{\ee}{\end{equation}}
\newcommand{\la}{\label}

\newcommand{\mm}{\mbox{\boldmath$g$}}
\expandafter\ifx\csname mathbbm\endcsname\relax

\else

\fi \textheight 22cm \textwidth 15cm \topmargin 1mm
\def\M{{\cal M}}

\def\Z{{\bf Z}_4}

\def\dzm{{\partial}}
\def\dzp{{\bar \partial}}

\def\a {{\alpha}}

\def\b {{\beta}}

\def\d {{\delta}}
\def\s {{\sigma}}

\def\mh {{\widehat\mu}}

\def\ad {{\dot\alpha}}

\def\bd {{\dot\beta}}

\def\mdh {{\widehat{\dot\mu}}}

\def\t {{\theta}}
\def\tb {{\bar\theta}}

\def\tba {{\bar\theta^\ad}}
\def\th {{\theta}}

\def\p {{\partial}}

\def\half {{\frac{1}{2}}}

\def\pp{{\mathchoice
          {
              \kern 1pt%
              \raise 1pt
              \vbox{\hrule width5pt height0pt depth0pt
                    \kern -2pt
                    \hbox{\kern 2.3pt
                          \vrule width0pt height6pt depth0pt
                          }
                    \kern -2pt
                    \hrule width5pt height0pt depth0pt}%
                    \kern 1pt
           }
            {
              \kern 1pt%
              \raise 1pt
              \vbox{\hrule width4.3pt height0pt depth0pt
                    \kern -1.8pt
                    \hbox{\kern 1.95pt
                          \vrule width0pt height5.4pt depth0pt
                          }
                    \kern -1.8pt
                    \hrule width4.3pt height0pt depth0pt}%
                    \kern 1pt
            }
            {
              \kern 0.5pt%
              \raise 1pt
              \vbox{\hrule width4.0pt height0.3pt depth0pt
                    \kern -1.9pt  
                    \hbox{\kern 1.85pt
                          \vrule width0.3pt height5.7pt depth0pt
                          }
                    \kern -1.9pt
                    \hrule width4.0pt height0.3pt depth0pt}%
                    \kern 0.5pt
            }
            {
              \kern 0.5pt%
              \raise 1pt
              \vbox{\hrule width3.6pt height0pt depth0pt
                    \kern -1.5pt
                    \hbox{\kern 1.65pt
                          \vrule width0pt height4.5pt depth0pt
                          }
                    \kern -1.5pt
                    \hrule width3.6pt height0pt depth0pt}%
                    \kern 0.5pt
            }
        }}

\def\mm{{\mathchoice
                       {
                             \kern 1pt
               \raise 1pt    \vbox{\hrule width5pt height0.4pt depth0pt
                                  \kern 2pt
                                  \hrule width5pt height0.4pt depth0pt}
                             \kern 1pt}
                       {
                            \kern 1pt
               \raise 1pt \vbox{\hrule width4.3pt height0.4pt depth0pt
                                  \kern 1.8pt
                                  \hrule width4.3pt height0.4pt depth0pt}
                             \kern 1pt}
                       {
                            \kern 0.5pt
               \raise 1pt
                            \vbox{\hrule width4.0pt height0.3pt depth0pt
                                  \kern 1.9pt
                                  \hrule width4.0pt height0.3pt depth0pt}
                            \kern 1pt}
                       {
                           \kern 0.5pt
             \raise 1pt  \vbox{\hrule width3.6pt height0.3pt depth0pt
                                  \kern 1.5pt
                                  \hrule width3.6pt height0.3pt depth0pt}
                           \kern 0.5pt}
                       }}

\begin{document}

\baselineskip 6mm

\begin{titlepage}
\hfill
\begin{flushright}
\end{flushright}

\vspace*{10mm}
\begin{center}
{\Large {\bf Black Hole Partition Function using Hybrid Formalism of Superstrings\\ }}
\vspace*{15mm} \vspace*{1mm} {B. Chandrasekhar
%
}\\
\vskip 0.5cm
{Instituto de F\'{i}sica Te\'{o}rica, Universidade Estadual Paulista, \\
Rua Pamplona, 145, 01405-900, S\~{a}o Paulo, Brasil}\\
\vskip 0.1cm
\end{center}

\vskip 1.0cm

\begin{abstract}

The type IIA superstring partition function $Z_{\rm IIA}$ on the euclidean attractor geometry $AdS_2 \times S^2 \times CY_3$, computes the modified elliptic genus $Z_{\rm BH}$ of the associated black hole. The hybrid formalism of superstrings defined as a conformally invariant sigma model on the coset supermanifold $PSU(1,1|2)/U(1)\times U(1)$, together with Calabi-Yau and chiral boson CFTs, is used to calculate $Z_{\rm IIA}$. The sigma model action on $AdS_2 \times S^2$ is explicitly written in $U(1)\times U(1)$ invariant variables. The $N=2$ generators of $AdS_2 \times S^2 \times CY_3$ are enlarged and embedded in an $N=4$ topological algebra. The world sheet superconformal invariance is then used to construct a nilpotent BRST operator, in contrast to the kappa symmetry analysis used by Beasely et. al. in hep-th/0608021. The sigma model action is explicitly shown to be closed under this BRST operator. Localization arguments are then used to deform the world sheet path integral with the addition of a BRST exact term, where, contributions arise only from the center of $AdS_2$ and, the north and south poles of $S^2$. This leads to the OSV result $Z_{\rm BH} = Z_{\rm IIA} = |Z_{\rm top}|^2$, where $|Z_{\rm top}|^2$ is the square of the topological string partition function.

\end{abstract}
\end{titlepage}
\section{Introduction} 
In~\cite{0405146}, OSV (Ooguri, Strominger and Vafa) conjectured a relationship of the form:
\be \label{osv}
Z_{\rm BH} = |Z_{\rm top}|^2
\ee
where $Z_{\rm BH} $ is the (indexed) entropy of four dimensional BPS black holes in Type II 
Calabi-Yau compactifications and $Z_{\rm top}$ is the square of topological string partition function 
evaluated at the attractor point on the associated Calabi-Yau. One way to view the relationship is to
think of it as an asymptotic expansion in the limit of large black hole charges. In this limit, 
$Z_{\rm BH}$ 
receives all order perturbative contributions from the F-term corrections in the low energy 
effective action of the ${\cal N} =2$ supergravity. For the Type IIA superstrings on Calabi-Yau three folds, 
these are of the general form:
\be \la{F}
\int d^4x d^4\theta ({ W}_{\a\b}{ W}^{\a\b})^g F_g(X^\Lambda).
\ee
where $X^\Lambda$ are the vector multiplet fields and the Weyl Superfield $W^{\a\b}$,
involves the graviphoton field strength and the Weyl tensor. These F-term corrections are
in turn captured by the topological string amplitudes $F_g$~\cite{Bershadsky:1993cx,Antoniadis:1993ze}. 
Together, with the fact that the
BPS black hole entropy gets corrected in the presence of these terms via attractor 
mechanism~\cite{Ferrara:1995ih}-\cite{Lopes Cardoso:1999ur},
and that the supergravity
partition function defines a mixed thermodynamic ensemble, provides the 
link (\ref{osv})~\cite{0405146}.  Several refinements of the relationship (\ref{osv}) have been suggested in literature~\cite{Vafa:2004qa}-\cite{Dijkgraaf:2000fq}. 

In particular, in~\cite{Gaiotto:2006ns}, an M-theory lift was used to calculate the black hole 
partition function at low temperatures as a dilute-gas sum over BPS wrapped 
membranes in $AdS_3\times S^2\times CY_3$, via the MSW CFT~\cite{Maldacena:1997de} and agrees with the
Gopakumar-Vafa partition function~\cite{Gopakumar:1998ii,Gopakumar:1998jq} (see related discussion 
in~\cite{deBoer:2006vg} ).  
The relation (\ref{osv}) is then obtained after
doing a modular transformation to go to high temperatures. Further, in~\cite{0608021} a 
computation of the black hole partition function without the need for a 
modular transformation was presented. 
In this method, first an M-theory lift of the IIA attractor geometry leads to 
a quotient of $AdS_3 \times S^2 \times CY_3$, whose asymptotic boundary is a
torus. Then, using $AdS_3$/$CFT_2$ duality, it is shown that the partition
function of IIA theory on the attractor geometry $AdS_2 \times S^2 \times {\rm CY}_3$, denoted as $Z_{\rm IIA}$, is equivalent to the partition function of the black hole. Starting from the euclidean Calabi-Yau
attractor geometry for the black hole, carrying D0-D2-D4 charges 
$q_0,q_A {(\rm electric)},p^A ({\rm magnetic})$ respectively, it was 
argued that~\cite{0608021}:
\be \label{zIIazbh}
{ Z_{IIA}(\phi^0,\phi^A,p^A) = Z_{BH}} = Tr_{R} (-)^Fe^{-{4\pi^2 \over
\phi^0} L_0 - q_A {\phi^A \over \phi^0}} 
\ee
where $\phi^0,\phi^A$ are potentials conjugate to the D0-D2 charges. 
The trace on the right hand side runs over the black hole microstates computed
in the CFT on the boundary torus dual to the M-theory lift of the 
IIA attractor geometry. In essence, the black hole partition function can be evaluated
in terms of the type IIA partition function 
on $AdS_2 \times S^2  \times CY_3$ geometry~\cite{Guica:2007wd}.


In~\cite{0608021}, the Green-Schwarz formalism of superstrings was used to
set up such a calculation, based on the general procedure for computing instanton generated
superpotential and higher derivative F-terms in the low energy effective action of string
theory~\cite{Beasley:2005iu}. The idea is to evaluate the IIA partition 
function on euclidean $AdS_2 \times S^2 \times {\mathcal M}$, in a perturbative string loop expansion,
where ${\mathcal M}$ is the moduli space of world-sheet instantons which wrap isolated holomorphic
curves inside the Calabi-Yau.
At genus $g$, the $AdS_2 \times S^2$ path integral receives contributions from genus 
g (anti)instantons which wrap
(anti)holomorphic cycles in the Calabi-Yau and sit at the center of $AdS_2$ and (north)south
poles of $S^2$. The instantons typically break some supersymmetries and one has to integrate 
over the resulting zero modes in the path integral. The analysis of~\cite{0608021} involved regularizing the
divergence coming from such integrals using 
$\kappa$-symmetry (see~\cite{Sen:2008yk}-\cite{Sen:2008vm} for a different approach and ~\cite{Sen:2009vz}-\cite{Dabholkar:2010uh} for recent work). Furthermore, the Calabi-Yau part of the instanton 
partition function was assumed to produce the topological string partition function 
and using some input from supergravity, the relation (\ref{osv}) was obtained.

The purpose of this paper is to use the hybrid formalism~\cite{9907200} to calculate the
partition function of IIA superstrings on $AdS_2 \times S^2$ with Ramond-Ramond flux. 
There are several advantages of this approach over the Green-Schwarz formalism. First, the
calculation of scattering amplitudes in hybrid formalism can be performed in a superpoincar\'{e} 
covariant manner and is relevant for describing $d=4, N=2$ theories. The calculation of
superspace low energy effective actions in Calabi-Yau compactifications and their 
connection to topological string amplitudes can be derived nicely~\cite{9510106,9407190}. 
Another important reason is that studying string propagation in Ramond-Ramond backgrounds is in general 
difficult in the traditional RNS formalism due to the need to introduce spin fields. In the Green-Schwarz formalism,
covariant quantization is problematic, except in light-cone gauge. 
To address these perennial issues, several covariant 
quantization techniques have been developed over the last few years by Berkovits and collaborators. In ten
dimensions, the pure spinor
formalism~\cite{Berkovits:2000fe} has led to a remarkable simplification of the calculation of superstrings loop amplitudes in flat space and their equivalence to the RNS formalism  has been fully demonstrated (see e.g.,~\cite{Berkovits:2001us,Berkovits:2004px,Berkovits:2004tw, Mafra:2005jh,Berkovits:2005ng} and the non-minimal approach~\cite{Berkovits:2005bt,Berkovits:2006bk,Berkovits:2006vc,Berkovits:2006vi}). 
The Formulation of the pure spinor superstring in $AdS_5 \times S^5$ background~\cite{Berkovits:2000fe,Berkovits:2000yr,Berkovits:2004xu}, following earlier studies on hybrid formalism in $AdS_3 \times S^3$~\cite{Berkovits:1999du,Berkovits:1999im} and in $AdS_2 \times S^2$~\cite{9907200} backgrounds, is being actively used in a world-sheet approach to the Maldacena conjecture~\cite{Berkovits:2007zk,Berkovits:2008qc,Berkovits:2007rj}. 

There have been innumerable applications of the hybrid formalism of superstrings in the presence of RR-backgrounds, to list a few, scattering amplitudes in lower dimensions~\cite{9407190,Berkovits:2001nv}, in obtaining C-deformation~\cite{Ooguri:2003qp} from superstrings in
graviphoton backgrounds~\cite{Berkovits:2003kj}, emergence of non(anti)-commutative superspace~\cite{Seiberg:2003yz,de Boer:2003dn}, 
for Calabi-Yau compactifications to two dimensions~\cite{Berkovits:2001tg} 
and more recently, in the study of flux vacua from the world-sheet point of view~\cite{Linch:2006ig,Linch:2008rw}.

We use the formulation of hybrid superstrings  
in the near horizon geometry of extremal black holes in four dimensions, presented in~\cite{9907200}. It was shown that, 
two-dimensional sigma models based on the coset supermanifold $\frac{PSU(1,1|2)}{U(1)\times U(1)}$ with Wess-Zumino term, provide
an elegant description of superstrings propagating on $AdS_2 \times S^2$ background with Ramond-Ramond flux. It is also
desirable to have a world-sheet perspective on our understanding of the
derivation of OSV relation (\ref{osv}); as mentioned earlier, the world-sheet approaches 
to large N topological string dualities are playing a special role in our understanding of the Maldacena 
conjecture~\cite{Berkovits:2007rj}. For the present case, the connection of low energy F-terms and the topological string amplitudes
can be understood nicely in the hybrid approach~\cite{9407190} (see~\cite{9604123,Kappeli:2006fj} for a review). More importantly,
in hybrid approach, the dilaton in type II theories, couples to the $N=(2,2)$ worldsheet supercurvature via the
Fradkin-Tseytlin term in the action~\cite{9510106}.
This has the advantage that the scattering amplitudes have a well defined 
dependence on the string coupling constant. This will be important while trying to get the right factors of topological
string coupling in the IIA partition function.

The rest of the paper is organized as follows. In Section-2, we review
the hybrid formalism of superstrings in flat space. In section-3, we 
discuss the $\frac{PSU(1,1|2)}{U(1)\times U(1)}$ sigma model and write down the
lowest order terms in the $AdS_2 \times S^2$ background in the $U(1)\times U(1)$ notation, using the expressions for left-invariant currents. In Section-4, we present various partition
functions in $AdS_2 \times S^2 \times CY_3$ related to the black hole partition function.
In Section-5, we present the computation of type IIA partition function over $AdS_2 \times S^2 \times CY_3$ by embedding the theory in $N=4$ topological strings. We provide localization
arguments using the BRST method and obtain the OSV relation.

\section{Hybrid Formalism of Superstrings}

The fundamental definition of $N=2$ superconformal theory describing hybrid
superstrings in four dimensions is by a field redefinition of $N=1$ RNS matter and 
ghost variables~\cite{9604123,Berkovits:1993zy,Berkovits:1993xq,ohta1,9404162,ohta2}. 
The $N=2, c=6$ system obtained after field redefinition is twisted and splits 
in to a four dimensional $c=-3, N=2$ superconformal field theory, coupled to an internal six dimensional $c=9, N=2$ superconformal theory. Now, one can proceed in two ways to
define physical states and calculate scattering amplitudes. The $c=6$ $N=2$ generators can be untwisted and coupled to a $c=-6, N=2$ ghost system to write down a BRST operator. Another way is to embed this $N=2$ system in small $N=4$ algebra and use the $N=4$ topological method, where there 
is no need to introduce ghost variables. The general four dimensional and six dimensional actions in arbitrary 
curved backgrounds with Ramond-Ramond flux, have been discussed in~\cite{9510106} and~\cite{Berkovits:1999du}, respectively. 
Below, we review the flat four dimensional case and consider the $AdS_2 \times S^2$ case in the next section.

\subsection{Flat Space-time}

For the case of type IIA superstrings compactified on Calabi-Yau 3-folds, the hybrid variables 
are as follows~\cite{9907200}. 
The four dimensional part consists of four bosonic and sixteen
fermionic spacetime variables given as: $X^m$ for $m=0$ to 3 
and $(\theta_{L}^\a ,
p_{L\a})$, $(\bar\theta_{L}^\ad, \bar p_{L\ad})$ $(\theta_{R}^\a,
p_{R\a})$, $({\bar\theta}_{R}{}^{\ad}, {\bar p}_{R\ad})$ where $\a,\ad = 1,2$.  
The six dimensional part has six bosonic 
variables $Y^j, \bar Y_j$ for $j=1$ to 3 and their associated 
superpartners, left-moving fermions $(\psi_L^j,\bar \psi_{Lj})$ and
right-moving fermions $(\psi_{R}^j,{\bar \psi}_{Rj})$ corresponding to the
Calabi-Yau. 
Finally, one has a chiral boson  $\rho_L$  and an anti-chiral boson $\rho_R$. 
The world sheet action corresponding to these fields in the superconformal
gauge can be divided into three parts as:
\be \label{4dcyrho}
 S_{\rm d=4} \, + \, S_{\rm CY} \, + \, S_{\rho_L,\rho_R} \, .
\ee
The four dimensional part of the action in flat superspace is:
\be \label{4da}
S_{\rm d=4} = \frac{1}{\a '}
\int dz d{\bar z}\,\, \Bigl[\,\half\dzp X^m \dzm X_m + p_{L\a} \dzp\t_L^\a +
\bar p_{L\ad} \dzp\tb_L^\ad + p_{R\a} \dzm \t_L^\a + {\bar p}_{R\ad} \dzm\tb_R^\ad \Bigr] ,
\ee

The Calabi-Yau part is:
\be
S_{\rm CY} = {\frac{1}{\a '}}
\int dz d \bar z \,\, \Bigl(\, \dzp Y^j \dzm \bar Y_j + \psi_L^j \dzp \bar\psi_{Lj}
+ \psi_R^j \dzp {\bar\psi}_{Rj} \,\Bigl) \, ,
\ee
and $ S_{\rho_L,\rho_R} $ stands for the action of the chiral and anti-chiral
bosons. For the present case, all the world sheet fields satisfy periodic 
boundary conditions. The free field OPE's corresponding to the above action are:
\bea
&&{X^m(y)X^n(z) \to - \alpha'\eta^{mn} \ln|y-z|^2, } \cr
&&p_{L\a}(y)\theta_L^\b (z)\to {\alpha'\delta_\a^\b\over{y -z}},\quad
\bar p_{L\ad}(y)\bar\theta_L^\bd (z)\to {\alpha'\delta_\ad^\bd\over{y -z}},\cr
&&p_{R\a}(y)\theta_R^\b (z)\to {\alpha'\delta_\a^\b\over{\bar y -\bar z}},\quad
\bar p_{R\ad}(y)\bar\theta_R^\bd (z)\to 
{\alpha'\delta_\ad^\bd\over{\bar y -\bar z}},\cr
&& \rho_L(y)\rho_L(z)\to -\ln(y-z),\quad 
\rho_R(y)\rho_R(z)\to -\ln(\bar y-\bar z)
\eea
The world sheet action (\ref{4dcyrho}) is manifestly conformally
invariant from the conformal dimensions of the fields. The action (\ref{4dcyrho}) further has an
$N = 2$ superconformal invariance realized nonlinearly, the left moving generators of which are\footnote{The superscripts $L,R$ on various generators stand for left and right-moving parts of the algebra. When these superscripts are not indicated explicitly, we always mean the left-moving part.
We also adopt the notation $|A|^2 = A_L\, A_R$}:
\bea \label{N2}
T^{L} &=&\half\dzm X^m  \dzm X_m +
p_{L\a}\dzm \t_{L}^\a + \bar p_{L\ad} \dzm\tb_{L}^\ad  + {{\alpha'}\over 2}
\dzm\rho_L\dzm\rho_L + T^{L}_{CY} \cr
G^{-L} &= & \frac{1}{\sqrt{\alpha'}}e^{\rho_L} (d_L\,)^2 
+ G^{-L}_{CY} \, \cr
G^{+L}  &=& \frac{1}{\sqrt{\alpha'}} e^{-\rho_L} ({\bar d}_L\,)^2
+ G^{+L}_{CY} \, \cr
J^L &= &  \alpha'\dzm\rho_L ~ 
+ J^{L}_{CY}
\eea
where,
\bea \label{d}
d_{L\a}=p_{L\a}+i\s^m_{\a\ad}\tb_{L}^{\ad}\dzm X_m -\half(\tb_L)^2\dzm\t_{L\a}
+{1\over 4}\t_{L\a} \dzm (\tb_L)^2, \cr
 \bar d_{L\ad}=\bar p_{L\ad}
+i\s^m_{\a\ad}\t_{L}^{\a}\dzm X_m -\half(\t_L)^2\dzm\tb_{L\ad}
+{1\over 4}\tb_{L\ad} \dzm (\t_L)^2~,
\eea
and $(d_L)^2$ means $\half\epsilon^{\a\b} d_{L\a} d_{L\b}$. The right-moving
$N=2$ generators are given as:
\bea \label{N2r}
T^{R} &=& \half\dzp X^m  \dzp X_m +
p_{R\a}\dzp \t_{R}^\a + \bar p_{R\ad} \dzp\tb_{R}^\ad  + {{\alpha'}\over 2}
\dzp\rho_R\dzp\rho_R + T^{R}_{CY} \cr
G^{-R} &= & \frac{1}{\sqrt{\alpha'}}e^{\rho_R} (d_R\,)^2 +  G^{+R}_{CY} \,, \cr
G^{+R}  &=& \frac{1}{\sqrt{\alpha'}}e^{-\rho_R} ({\bar d}_R\,)^2 
+ G^{-R}_{CY} \,, \cr
J^R &= & \alpha' \dzp\rho_R - J^{R}_{CY}
\eea
with,
\bea \label{dr}
d_{R\a}=p_{R\a}+i\s^m_{\a\ad}\tb_{R}^{\ad}\dzp X_m -\half(\tb_R)^2\dzp\t_{R\a}
+{1\over 4}\t_{R\a} \dzp (\tb_R)^2, \cr
 \bar d_{R\ad}=\bar p_{R\ad}
+i\s^m_{\a\ad}\t_{R}^{\a}\dzp X_m -\half(\t_R)^2\dzp\tb_{R\ad}
+{1\over 4}\tb_{R\ad} \dzp (\t_R)^2~,
\eea
The $\pm$ signs on the fermionic generators $G^{L,R}$ 
represent their $U(1)$ charges with respect to $J^{L,R}$. 
Also, $[T^{L,R}_{CY},G^{L,R\pm}_{CY}, G^{L,R\pm}_{CY},J^{L,R}_{CY}]$ are the left and right-moving, $N=2$ $c=9$ generators
describing the Calabi-Yau 3-fold. We have written the left and right-moving algebras explicitly for the type IIA superstring. The type IIB generators can be obtained by noting that under mirror symmetry, 
$J^R_{CY} \rightarrow - J^R_{CY}$, resulting in interchange  $G^{+R}_{CY} \rightarrow G^{-R}_{CY}$. The space-time part remains unaffected by this change.

The set of operators $d_{L\a},\bar d_{L\ad}$,
as shown in~\cite{Siegel:1985xj}, satisfy the algebra:
\be
{d_{L\a} (y) \bar d_{L\ad}(z) \to 2i\alpha'{{\Pi_{L\a\ad}}\over{y -z}}, \quad
d_{L\a}(y) d_{L\b}(z) \to {\rm reg.},
\bar d_{L\ad}(y) \bar d_{L\bd}(z) \to {\rm reg.},} 
\ee
with similar expressions for the right-moving part. Here,
\be \la{pi}
\Pi^m_L =\dzm X^m - \frac{i}{2}\s^m_{\a\ad}(\th_L^\a \dzm \tb_L^\ad+
\tb_L^\ad\dzm \th_L^\a) \, ,
\Pi^m_R =\dzp X^m - \frac{i}{2}\s^m_{\a\ad}(\th_R^\a \dzp \tb_R^\ad+
\tb_R^\ad\dzp \th_R^\a) \, ,
\ee
The operators $d_{L\a},\bar d_{L\ad}$ also commute with the spacetime supersymmetry generators of the action:
\bea \label{q}
q_{L\a} =\int dz \,Q_{L\a}~ ~~{\rm where}~~~Q_{L\a}=p_{L\a} -i\s^m_{\a\ad}\tba_{L}\dzm
X_m-
        {1\over 4}(\tb_L)^2\dzm\t_{L\a}~,\cr
\bar q_{L\ad}=\int dz \, \bar Q_{L\ad}~~~{\rm where}~~~
 \bar Q_{\ad}= \bar p_{L\ad}
        - i\s^m_{\a\ad}\t_{L\a}\dzm X_m-{1\over 4}(\t_L)^2\dzm\tb_{L\ad}~.
\eea
Further, $q_{R\a}$ and ${\bar q}_{R\ad}$ are obtained by 
replacing $L$ with  $R $ and $\partial$ with $\bar
\partial$. Thus, eqns. (\ref{d}) and (\ref{q}) contain the world-sheet versions of
the space-time superspace covariant derivatives and supersymmetry generators, 
respectively. 

\section{ Hybrid Formalism in $AdS_2 \times S^2$ }
The four dimensional part of the action $S_{\rm d=4}$
in eqn. (\ref{4dcyrho}), can be replaced with the world sheet action for the sigma model 
on $AdS_2 \times S^2$. Below, we discuss the $AdS_2 \times S^2$  model in detail. 
In~\cite{9907200} it was shown that the two-dimensional sigma models based
on the coset supermanifold $PSU(1,1|2)/U(1)\times U(1)$ can be used to 
quantize superstrings on $AdS_2 \times S^2$ in RR-backgrounds. One of the
main advantages of the hybrid approach is that 4D super-poincar\'{e} invariance and the
$N=2$ target space-time supersymmetry can be made manifest even in the presence of 
RR-fields. For instance, like in
Green-Schwarz formalism, the model has fermions which are spinors in target
space and scalars on the world-sheet allowing a simple treatment of RR-fields.
Second, like in RNS formalism, the action is quadratic in flat-space allowing
for quantization. As mentioned earlier, a first principle definition of the world-sheet 
variables in hybrid formalism is through a redefinition of RNS world-sheet
variables. This follows from the fact that any critical $N=1$ string can be embedded
in an $N=2$ string~\cite{Berkovits:1993xq}.

For type II superstrings, the four
$\kappa$-symmetries of the Green-Schwarz action are replaced by a critical world-sheet $N=(2,2)$ 
superconformal invariance. These local $N=2$ superconformal generators are nothing
but the twisted RNS world-sheet generators. The important
difference with RNS approach is that the vertex operators for RR-fields do
not have square-root cuts with supersymmetry generators and there is no need
to sum over spin structures on the world-sheet~\cite{9604123}.


\subsection{$AdS_2 \times S^2$ Sigma Model}

The $AdS_2 \times S^2$ sigma model action is constructed as a gauged principle 
chiral field~\cite{Bershadsky:1999hk,9907200}. 
Let {\fontfamily{ptm}\selectfont g}$(x) \in G $ denote a map from the worldsheet into the supergroup $G$, with
the current $J = ${\fontfamily{ptm}\selectfont g}$^{-1}d${\fontfamily{ptm}\selectfont g} valued in the Lie algebra ${\cal G}$. The 
sigma model action on the coset space is constructed by gauging a subgroup $H$,
whose Lie algebra is ${\cal H }_0$. 
To construct a string theory on $AdS_2 \times S^2$, one has to quotient the
group $ G \equiv PSU(1,1|2)$ by (the right action of) bosonic 
subgroups $ H \equiv U(1) \times U(1)$. This subgroup is in fact the invariant locus 
of a $\Z$ automorphism, the existence of which, plays a crucial role in the construction. 
In effect, a $\Z$ decomposition of the  $ PSU(1,1|2)$ Lie
algebra ${\cal G}$ can be written as~\cite{9907200}:
\be
{\cal G}=  {\cal H}_0 \oplus {\cal H}_1 \oplus {\cal H}_2 \oplus {\cal H}_3~ , \nonumber
\ee
where the subspace ${\cal H}_k$ is the eigenspace of the ${\bf Z}_4$ generator
$\Omega$ with eigenvalue $(i)^k$.  The subspaces $ {\cal H}_1$ and $ {\cal H}_3$ contain
all the fermionic generators of the algebra ${\cal G}$ and ${\cal H}_0, {\cal H}_2$ contain
the bosonic generators of ${\cal G}$.

The $AdS_2 \times S^2$ sigma model action is a sum of the action on the coset supermanifold 
$PSU(1,1|2)/U(1)\times U(1)$ and a Wess-Zumino term and is given as\footnote{We use the notation, $J \equiv J_z, \bar J \equiv J_{\bar z}$.}~\cite{9907200}:
\be \label{coset}
S_{AdS_2 \times S^2}={1 \over \pi \lambda^2 } \int d^2 x~
Str \left( \frac{1}{2} J_{2} {\bar J}_{2} +
\frac{3}{4} J_{1} {\bar J}_{3} +
\frac{1}{4} {\bar J}_{1}  J_{3}\right) .
\ee
where the left-invariant currents are given as,
\bea
J_{2}  &=& ({\rm g}^{-1}\p {\rm g})^{m} T_{m}\, , \cr
J_{1} &=& ({\rm g}^{-1}\p {\rm g})_{R}^{\a} T^{R}_{\a} +  ({\rm g}^{-1}\p {\rm g})_{R}^{\ad} T^{R}_{\ad},\cr
J_{3} &=& ({\rm g}^{-1}\p {\rm g})_{L}^{\a} T^{L}_{\a} +  ({\rm g}^{-1}\p {\rm g})_{L}^{\ad} T^{L}_{\ad},
\eea
where $T_m,T^{R,L}_{\a,\ad}$ denote the $PSU(1,1|2)$ generators corresponding to translations and
supersymmetry. ${\bar J}$'s can be obtained by replacing $\p$ with $\bar \p$ and 
$Str$ stands for supertrace over the $PSU(1,1|2)$ matrices,
the algebra of which is given in eqn. (\ref{psu}). The action in 
eqn. (\ref{coset}) has invariance under global $PSU(1,1|2)$ transformations
which are realized by left-multiplication as $\delta {\rm g} = (\Sigma^A T_A) {\rm g}$.
There is also an invariance under local $U(1) \times U(1)$ gauge transformations, which is
realized on ${\rm g}(x,\theta)$ by a right multiplication as $\delta {\rm g}= {\rm g}\Sigma(x)$. The supersymmetry transformations (up to a local Lorentz rotation) 
of the superspace coordinates are achieved by a global left action of the superalgebra on the coset
superspace, where as the $\kappa$-supersymmetry corresponds to a local right
action on the same coset superspace. The $\kappa$-supersymmetry is explicitly broken in the present
case and is replaced by world-sheet superconformal invariance~\cite{Berkovits:2002zk}.

It is useful to write the action by introducing auxiliary
fields as:
\bea \la{actiond}
S_{AdS_2 \times S^2} &=& \frac{1}{\alpha'} \int d^2z \left[ \; \frac{1}{2} \eta_{cd} J^c{\bar J}^d 
- \frac{1}{4Ng_s}\delta_{\alpha_L\beta_R}\,\left( J^{\alpha_L}{\bar J}^{\beta_R}
- {\bar J}^{\alpha_L} J^{\beta_R} \right) \right. \cr
&-& \left. \frac{1}{4Ng_s}\delta_{\dot\alpha_L\dot\beta_R}\,\left( J^{\dot\alpha_L}{\bar J}^{\dot\beta_R}
- {\bar J}^{\dot\alpha_L} J^{\dot\beta_R} \right) 
+ d_{\alpha_L} {\bar J}^{\alpha_L} + 
{\bar d}_{\dot\alpha_L} {\bar J}^{\dot\alpha_L} + d_{\alpha_R} {J}^{\alpha_R} + 
{\bar d}_{\dot\alpha_R} {J}^{\dot\alpha_R} \right. \cr
&+& \left. Ng_sd_{\alpha_L}d_{\beta_R} \delta^{\alpha_L\beta_R}
+ Ng_s {\bar d}_{\dot\alpha_L} {\bar d}_{\dot\beta_R} \delta^{\dot\alpha_L\dot\beta_R} \, 
\right]\, .
\eea
This form of the action can be obtained by writing the flat space-time action in (\ref{4da}), in terms
of manifestly supersymmetric variables of eqns. (\ref{pi}) 
and (\ref{jpirelation}), and then
covariantizing in a curved background~\cite{9907200}.
Using the following constraint equations,
\be \label{dconstraint}
d_{\alpha_L} = \frac{1}{Ng_s}\, \delta_{\alpha_L\beta_R}\, J^{\beta_R}, \qquad
d_{\alpha_R} = -\frac{1}{Ng_s}\, \delta_{\beta_L\alpha_R}\, {\bar J}^{\beta_L},
\ee
and performing the scaling,
\be
E^c_M \rightarrow (Ng_s)^{-1}E^c_M, \quad 
E_M^{\alpha_{L,R}}, E_M^{\dot\alpha_{L,R}} \rightarrow (Ng_s)^{-\frac{1}{2}}E_M^{\alpha_{L,R}}, E_M^{\dot\alpha_{L,R}}
\ee
one arrives at the action:
\bea \label{cosetA}
S_{AdS_2 \times S^2} &=& \frac{1}{\alpha'N^2g_s^2} \int d^2z \left[ \; \frac{1}{2} \eta_{cd} J^c{\bar J}^d 
- \frac{1}{4}\delta_{\alpha_L\beta_R}\,\left( J^{\alpha_L}{\bar J}^{\beta_R}
+ 3 {\bar J}^{\alpha_L} J^{\beta_R} \right) \right. \cr
&-& \left. \frac{1}{4}\delta_{\dot\alpha_L\dot\beta_R}\,\left( J^{\dot\alpha_L}{\bar J}^{\dot\beta_R}
+3 {\bar J}^{\dot\alpha_L} J^{\dot\beta_R} \right)  \right]\, .
\eea

The above $PSU(1,1|2)$ invariant action can be written in a more familiar
form for type $II$A superstrings on $AdS_2 \times S^2 $ with Ramond-Ramond fluxes 
in a manifestly space-time supersymmetric manner using the following
identifications~\cite{9907200},
\be \label{jpirelation}
J^{c} \equiv \Pi^{c} , \qquad
J^{\alpha_{L,R} } \equiv  {\Pi}{}^{\alpha_{L,R}} , \qquad 
J^{\dot\alpha_{L,R}} \equiv  {\Pi}{}^{\dot\alpha_{L,R}}  \, ,
\ee
and similarly for $\bar J$'s as well.
Here, $\Pi^A_j=E_M{}^A \partial _j Z^M$, $Z^M=(X^m,
\t_L^\mu,\tb_L^{\dot\mu},\t_R^\mh, \tb_R^\mdh)$, supervierbein $E_M {}^A$  
and $A$ takes the tangent-superspace
values [$c,\alpha_L,\ad_L,\a_R,\ad_R$]. The full action is now:
\be \label{adscyrho}
S = S_{AdS_2 \times S^2} \, + \, S_{\rm CY} \, + \, S_{\rho_L,\rho_R} \, .
\ee

We now collect the $N=2$ generators
in this notation:
\bea \la{ads2s2N2}
T &=&  T_{{\rm d}=4}  + T_{CY}~,\cr
G^- &=& G^-_{{\rm d}=4}~+ G^-_{CY}, \cr
G^+ &=& G^+_{{\rm d}=4}~+ G^+_{CY}, \cr
J &=& J_{{\rm d}=4}~+ J_{CY}~,
\eea
where ,
\bea \la{4dalgebra}
T_{{\rm d}=4} &=&  \Pi^c_{\bar z} \Pi_{c{\bar z}} + d_\a \Pi^\a_{\bar z} +  {\bar d}_\ad 
\Pi^{\ad}_{\bar z}
 +{1\over 2}
\dzp \rho \dzp \rho, \cr
G^-_{{\rm d}=4} &=& e^{\rho} (d)^2 ,\cr
G^+_{{\rm d}=4} &=& e^{-\rho}  (\bar d)^2 \, ,\cr
J_{{\rm d}=4} &=& \dzp \rho \, ,
\eea
represent the generators of $c=-3, N=2$ algebra describing $AdS_2 \times S^2$, and
\bea \label{N2cy}
T_{CY} &=&\dzp Y^j\dzp \bar Y_j +\half(\psi^j\dzp\bar \psi_{j}
+\bar\psi_{j}\dzp \psi^j)~, \cr
G^{-}_{CY} &= & \psi^j\dzp \bar Y_j\,, \cr
G^{+}_{CY}  &=&  +\bar \psi_{j} \dzp Y^j,\cr
J_{CY} &= &  \psi^j\bar\psi_{j}~,
\eea 
represent the $c=9, N=2$ generators describing the Calabi-Yau 3-fold. The generators of
$AdS_2 \times S^2$ algebra, given in eqn. (\ref{4dalgebra}), are similar to the ones
of flat-space time given eqns. (\ref{N2}) and (\ref{N2r}), except for the constraints satisfied by $d$ in eqn. (\ref{dconstraint}). One can use the equations of
motion in (\ref{dconstraint}), to write the four dimensional
generators explicitly in terms of the left-invariant currents.  
We also note that the fermionic generators in (\ref{4dalgebra}), remain
holomorphic even at the quantum level~\cite{9907200}. 
The generators in eqn. (\ref{ads2s2N2}) correspond to a $c=6,N=2$ algebra. In section-\ref{variousP}, we topologically twist this algebra and embed it in an $N=4$ algebra.

\subsection{$AdS_2 \times S^2 $ Action in $U(1) \times U(1)$ 
Notation} \label{u1notation}

For the localization arguments to be presented in the section-\ref{localization}, it will further be useful to write the action (\ref{cosetA}) in manner in which the $U(1) \times U(1)$ charges
are manifest. This notation is quite useful, as most of the
algebraic relations can be easily obtained just from constructing $U(1) \times U(1)$ invariant expressions. The sigma model on the coset space $G/H$ with the Wess-Zumino
term is one-loop conformally invariant provided that the group G is Ricci flat and H is the
fixed locus of a $Z_4$ automorphism of G. For instance, $J_{1}$ decomposes under $H=U(1) \times U(1)$ as\footnote{The fermionic generators have appropriate flat space limits; for 
instance, $J_{++,--} \rightarrow d_{\alpha}$,
$J_{+-,-+} \rightarrow {\bar d}_{{\dot\alpha}}$. We take the $++,--$ subscripts to follow from undotted indices of fermions and $+-,-+$ indices from dotted ones.}:
\be \label{z4J}
J_{1} \equiv J_R^{++}, J_R^{--}, J_R^{-+}, J_R^{+-}
\ee 
Now, in terms of the decomposition given in (\ref{z4J}), the fermionic generators of the
$N=2$ superconformal algebra in  $AdS_2 \times S^2$ can be constructed as~\cite{9907200}:
\be \label{gl}
G^{-R}_{{\rm d}=4} =  \int dz \left( e^{\rho_R} \,J_R^{++}J_R^{--} \right), \qquad
G^{+R}_{{\rm d}=4}  = \int dz \left( e^{-\rho_R} \,J_R^{+-}J_R^{-+} \right)
\ee
and also the left-moving part involving $J^{3}$. The transformations of various fields 
under these generators are obtained by a global right multiplication of g. Note that unlike in flat space, the
currents have dependence on both holomorphic and anti-holomorphic coordinates.

Let us start by writing down the $PSU(1,1|2)$ algebra relations.
The generators of the bosonic subgroup $SO(2,1) \times SO(3)$ are $M, T_{\pp 0},T_{\mm 0} $,
$N, T_{0\pp},T_{0\mm}$, where the subscripts denotes charges under $U(1)\times U(1)$. The fermionic
generators are $T^i_{++},T^i_{--}$ and $T^i_{+-},T^i_{-+}$ where $i,j = L,R$. Thus, the relevant relations of
the $PSU(1,1|2)$ algebra are:
\bea \label{psu}
&&\left[ T_{\pp 0}, T_{\mm0} \right] = M, \qquad \left[ T_{0\pp }, T_{0\mm} \right] = N, \qquad
\{T^{i}_{++},T^{j}_{+-} \} = \delta^{ij}T_{\pp 0}, \cr
&&\{T^{i}_{++},T^{j}_{-+} \} = \delta^{ij}T_{0\pp}, \qquad
\{T^{i}_{--},T^{j}_{+-} \} = \delta^{ij}T_{0\mm},\qquad \{T^{i}_{--},T^{j}_{-+} \} = \delta^{ij}T_{\mm0}, \cr
&&\{T^{i}_{++},T^{j}_{--} \} = \epsilon^{ij} \left(N-M \right), \qquad
\{T^{i}_{-+},T^{j}_{+-} \} = \epsilon^{ij} \left( N+M \right) \, ,
\eea
where $\epsilon^{LR} = 1 = - \epsilon^{RL}$. Using the above generators, the action in (\ref{cosetA}) can be written as:
\bea \label{cosetU}
S_{AdS_2 \times S^2} &=& \frac{1}{\alpha'N^2g^2} \int d^2z \left[ \; \frac{1}{2} ( J^{\pp 0}{\bar J}^{\mm 0} - J^{0\pp}{\bar J}^{0\mm} 
+ J^{\mm 0}{\bar J}^{\pp 0} - J^{0\mm}{\bar J}^{0\pp} ) \right. \cr
&-& \left. \frac{1}{4}
\left( J_L^{--}{\bar J}_R^{++}
+ 3 {\bar J}_L^{--} J_R^{++} + J_L^{++}{\bar J}_R^{--}
+ 3 {\bar J}_L^{++} J_R^{--} \right) \right. \cr
&-& \left. \frac{1}{4}
\left( J_L^{+-}{\bar J}_R^{-+}
+ 3 {\bar J}_L^{+-} J_R^{-+} + J_L^{-+}{\bar J}_R^{+-}
+ 3 {\bar J}_L^{-+} J_R^{+-}\right)  \right]
\eea
This action and its equations of motion are important to the localization arguments to be presented in section-\ref{localization}. Using the variation of a general current $J$ as $\delta J = d \delta X + [J, \delta X]$, we obtain the equations of motion of fermionic currents, following from (\ref{cosetU}) as,
\bea \la{eomads2s2}
&& \nabla {\bar J}_L^{--} = 0 \, , \qquad
\nabla {\bar J}_L^{++} = 0 \, , \qquad
\nabla {\bar J}_L^{-+} = 0 \, , \qquad
\nabla {\bar J}_L^{+-} = 0 \, , \cr
&& {\bar \nabla} J_R^{--} = 0 \, , \qquad
{\bar \nabla} J_R^{++} = 0 \, , \qquad
{\bar \nabla} J_R^{-+} = 0 \, , \qquad
{\bar \nabla} J_R^{+-} = 0 \, , 
\eea
where we have only written the equations for covariantly holomorphic and anti-holomorphic currents. Here, $\nabla J= \partial J + [J_0, J] $ with $J_0$ denoting the generators in the 
subspace ${\cal H}_0$ stands for the covariant derivative,  and similarly for $\bar \nabla$.
Using the parameterization of the group element as ${\rm g} = \exp(XT + \theta_LT_L + \theta_R T_R)$,
the left-invariant currents are:
\bea \label{Jone}
J^{\pp 0} &=& \partial X^{\pp 0} + \frac{1}{2}\left(\partial \theta_L^{++}\theta_L^{+-} + \partial \theta_L^{+-}\theta_L^{++} 
+ L \leftrightarrow R  \right ) + \cdots, \cr
J^{\mm 0} &=& \partial X^{\mm 0} + \frac{1}{2}\left(\partial \theta_L^{--}\theta_L^{-+} + \partial\theta_L^{-+}\theta_L^{--} 
+ L \leftrightarrow R  \right ) + \cdots , \cr
J^{0\pp} &=& \partial X^{0\pp} + \frac{1}{2}\left(\partial \theta_L^{++}\theta_L^{-+} + \partial \theta_L^{-+}\theta_L^{++} 
+ L \leftrightarrow R  \right ) + \cdots, \cr
J^{0\mm} &=& \partial X^{0\mm } + \frac{1}{2}\left(\partial \theta_L^{--}\theta_L^{+-} + \partial\theta_L^{+-}\theta_L^{--} 
+ L \leftrightarrow R  \right ) + \cdots ,
\eea
and the barred ones can be obtained similarly. The left-handed currents for instance are:
\bea \label{Jtwo}
J_L^{++} &=&  \partial\theta_L^{++} + \frac{1}{2} \left( -\partial X^{\pp 0}\theta_R^{-+} + \partial\theta_R^{-+}X^{\pp 0} +\partial X^{0\pp }\theta_R^{+-} - \partial\theta_R^{+-}X^{0\pp} \right) + \cdots , \cr
J_L^{--} &=&  \partial\theta_L^{--} + \frac{1}{2} \left( -\partial X^{\mm 0}\theta_R^{+-} + \partial\theta_R^{+-}X^{\mm 0} +\partial X^{0\mm }\theta_R^{-+} - \partial\theta_R^{-+}X^{0\mm} \right) + \cdots , \cr
J_L^{+-} &=&  \partial\theta_L^{+-} + \frac{1}{2} \left( \partial X^{\pp 0}\theta_R^{--} - \partial\theta_R^{--}X^{\pp 0} -\partial X^{0\mm }\theta_R^{++} + \partial\theta_R^{++}X^{0\mm} \right) + \cdots , \cr
J_L^{-+} &=&  \partial\theta_L^{-+} + \frac{1}{2} \left(\partial X^{\mm 0}\theta_R^{++} - \partial\theta_R^{++}X^{\mm 0} -\partial X^{0\pp }\theta_R^{--} + \partial\theta_R^{--}X^{0\pp} \right) + \cdots ,
\eea
and the right-handed currents and barred ones can be obtained similarly. Thus, the first few terms relevant for us in the action (\ref{cosetU}) are:
\bea \label{ads2s2}
S &=&   \int d^2z \left( \partial X^{\pp 0} \bar\partial X^{\mm 0}  -  \partial X^{0\pp} \bar\partial X^{0\mm}
-   \partial\theta_L^{++} {\bar\partial}\theta_R^{--} -  \partial \theta_L^{--} {\bar\partial}\theta_R^{++} \right. \cr
&+& \left. \partial\theta_L^{+-}{\bar\partial}\theta_R^{-+} + \partial\theta_L^{-+}{\bar\partial}\theta_R^{+-} + \cdots \right)
\eea
The OPE's corresponding to the action (\ref{ads2s2}) are:
\bea \label{opes}
&&{X^{\pp 0}(y) X^{\mm 0}(z) \sim -\ln|y-z|, \qquad X^{0\pp}(y) X^{0\mm}(z) \sim + \ln|y-z|} ,\cr
&&  \theta_L^{++}(y)\theta_R^{--}(z) \sim \ln|y-z|, \qquad  \theta_L^{--}(y)\theta_R^{++}(z) \sim \ln|y-z| ,\cr
&&  \theta_L^{+-}(y)\theta_R^{-+}(z) \sim -\ln|y-z|, \qquad  \theta_L^{-+}(y)\theta_R^{+-}(z) \sim -\ln|y-z| \, .
\eea
Notice that the auxiliary fields have been integrated out in the action in (\ref{ads2s2}). There are new OPE's among the fermionic
coordinates in the $AdS_2 \times S^2$ background, as in the $AdS_3 \times S^3$ case~\cite{Berkovits:1999im}.

\section{Black Hole and Topological String Partition Functions} \label{variousP}

In this section, we give details of the black hole and topological string partition functions to be used in the next section. 
We start with the topological string amplitudes of $N=2$ and $N=4$ topological strings. The partition function of $N=2$ topological strings leads to the right
hand side of relation in eqn. (\ref{osv}) and that of $N=4$ topological strings will 
lead to a method of calculating the partition function of the physical type IIA superstrings, appearing in eqn. (\ref{zIIazbh}). In the final subsection, we present the black hole partition function.

\subsection{$N=2$ Topological String Partition Function}\label{N2topstrings}
To define the $N=2$ topological strings, we use the generators of two copies of the $c=9,{\mathcal N}=2$ algebra of the Calabi-Yau, given in eqn. (\ref{N2cy}): $T^L_{\rm CY},G^{-L}_{\rm CY}, G^{+L}_{\rm CY}, J^L_{\rm CY} $ and 
$T^R_{\rm CY},G^{-R}_{\rm CY}, G^{+R}_{\rm CY}, J^R_{\rm CY} $, where all the fermionic generators $G^{L,R\pm}_{\rm CY}$ have spin $\frac{3}{2}$. Now, we perform a twist which shifts the $U(1)$ charges of the fields, leading to a topological theory. An A-twist corresponds to the choice of shifts, $T^L_{\rm new,CY} = T^L_{\rm CY} - \frac{1}{2}\dzp J^L_{\rm CY}$ and $T^R_{\rm new,CY} = T^R_{\rm CY} + \frac{1}{2}\dzm J^R_{\rm CY}$. This results in the
shift of spins $h^L,h^R$ of the generators of left and right-moving algebras as, $h^L_{\rm new} = h^L - \frac{1}{2}q^L$ and $h^R_{\rm new} = h^R + \frac{1}{2}q^R$, where $q^L,q^R$ 
are the $U(1)$ charges, respectively. Note that $T_{\rm new,CY}(y)\,T_{\rm new,CY}(z)$ OPE's have no central charge in the twisted $N=2$ system, and all bosonic and fermionic worldsheet fields have integer spin. However, $G^+_{\rm CY}(y)\,G^-_{\rm CY}(z)$ OPE's have a central charge given as $\hat c= 3$~\cite{Bershadsky:1993cx}. As a result of this twisting, $G^{-L}_{\rm CY}, G^{+R}_{\rm CY}$ acquire spin 2 and  $G^{+L}_{\rm CY}, G^{-R}_{\rm CY}$ acquire spin 1. The
$N=2$ topological string amplitude is then given as~\cite{Bershadsky:1993cx},
\begin{equation} \label{Fg} 
F_g = \int_{\M_g} \IP{\abs{\prod_{i=1}^{3g-3} G^-_{\rm CY}(\mu_i)}^2}\, ,
\end{equation}
where, $<|G^-_{\rm CY}(\mu)|^2>$ corresponds to $<G^{-L}_{\rm CY}(\mu)\,G^{+R}_{\rm CY}(\mu)>$. 
The spin 2 generators $G^{-L}_{\rm CY}$ and $G^{+R}_{\rm CY}$ appear in the amplitude, and behave similar to the b-ghost in the bosonic string. The spin 1 generators can be used to form a fermionic nilpotent operator, to study the cohomology of the theory. The formula (\ref{Fg}) should also be understood as coming from
coupling the $N=2$ theory to topological gravity. One can define the full topological string free energy to be
\begin{equation} \label{FreeEnergy}
{\mathcal F}_{top} = \sum_{g=0}^\infty g_{top}^{2g-2} F_g,
\end{equation}
where $g_{top}$ is the coupling constant weighing the contributions at
different genera. Finally, the topological 
string partition function is
defined as
\begin{equation} \label{ztop1}
Z_{\rm top} = \exp {{\mathcal F}_{top}}.
\end{equation}
One can repeat the above discussion for an $\bar A$-twist, corresponding to the choice of shifts, $T^L_{\rm new} = T^L + \frac{1}{2}\dzp J^L$ and $T^R_{\rm new} = T^R - \frac{1}{2}\dzm J^R$, leading up to the anti-topological
string partition function $\bar Z_{\rm top}$. $|Z_{\rm top}|^2$ in eqn. (\ref{osv}) then stands for the product of the $N=2$ topological and its conjugate, the anti-topological string partition functions, $Z_{\rm top}$ and ${\bar Z}_{\rm top}$, respectively.

\subsection{$N=4$ Topological Strings}\label{N4topstrings}

The $N=2$ topological strings discussed in the previous subsection, were modeled after the $N=0$ bosonic strings.  In a similar vein, $N=4$ topological strings consisting of a twisted small $N=4$ algebra, are modeled after $N=2$ strings.  The twisted ${\cal N}=4$ generators
are,
\be \label{N4fullgen}
\{~T,G^+,\tilde G^+, G^-,\tilde G^-, J^{++},J, J^{--}~\} \, \,
\ee
and correspond to an algebra with central charge $\hat c = 2$~\cite{9407190}.
The OPE's of the
algebra are given in~\cite{9407190} and one sees that there are two doublets $(G^+, \tilde G^-)$ and  $(\tilde G^+, G^-)$.
There is also an $SU(2)_f$ flavor which rotates these doublets and leaves the $N=4$ unchanged. Explicitly, the flavor rotation
is:
\bea \label{su2f}
&& \widehat{{\tilde G}^+}(u)=u_1 {\tilde G}^+ +u_2 { G^+}, \quad \widehat {G^-}(u)=u_1 { G}^--u_2{\tilde G^-} \cr
&& \widehat {{\tilde G}^-}(u)=u_2^*{\tilde G}^- - u_1^*{ G^-}, \quad \widehat {G^+}(u) =u_2^*{G}^+ +u_1^* {\tilde G^+} \, ,
\eea
where, $|u_1|^2+|u_2|^2=1$ and the complex conjugate of $u_a$ is $\epsilon^{ab} u^*_b$~\cite{Galperin:1984av}. The generators
${\tilde G}^{\pm}$, appearing in eqn. (\ref{su2f}), are defined as:
\be \label{Gtilde}
\tilde G^- \equiv \left[\oint J^{--}, G^+\right] ,\quad
\tilde G^+ \equiv \left[\oint J^{++}, G^-\right],
\ee 
where,
\be
J^{++}\equiv \exp \left(\int^z J\right),
\quad\quad J^{--}\equiv \exp \left(-\int^z J\right).
\ee

\noindent
The flavor rotated  $N=4$ topological string amplitude is given by
~\cite{9407190}:
\bea \label{t4}
Z&=&\sum_{n=-2g+2}^{2g-2}{(4g-4)!\over (2g-2+n)!(2g-2-n)!}
F_g^n u_1^{2g-2+n} u_2^{2g-2-n} \cr
&=&\int_{{\cal M}_g}\langle |\widehat {G^-}(\mu_1)...\widehat {G^-}(\mu_{3h-3})|^2\big[
\int \widehat{{\tilde G}^+} \widehat{\overline {\tilde
G^+}} \big]^{g-1}\int J\bar J\rangle
\eea
where $\mu's$ denote the Beltrami differentials and the $\int J$ type of insertions are needed 
to ensure that the path integral does not vanish trivially. Also, the $\int \widehat{\tilde G^+}$ insertions
in the path integral allow the use of $N=2$ topological strings in the calculation of amplitudes. 

\noindent
Now, one constructs unintegrated and integrated vertex operators and calculates scattering amplitudes for any critical $N=2$ strings. A case relevant for the present situation is the scattering of $2g-2$ chiral graviphotons and 
two gravitons in flat four dimensional space-time. This calculation was performed in~\cite{9407190}
and contributes terms of the kind in eqn. (\ref{F}) for the low energy effective action of type II
strings~\cite{Antoniadis:1993ze,Bershadsky:1993cx}. 
In~\cite{Beasley:2005iu} (see also~\cite{Antoniadis:1994hg,Antoniadis:1996qg,Antoniadis:2005sd}), a physical gauge formalism was used to study in a general
situation, what kind of terms in the
low energy effective action of superstrings can be generated by world sheet instantons which
wrap holomorphic curves $C$ in the Calabi-Yau. The A-model computation consists
of counting the bosonic and fermionic zero modes contributing to the instanton partition function.
In this case, the instanton that contributes is isolated and  of genus $g \ge 1$. One also needs to 
bring down appropriate number of insertions of interaction terms 
involving Weyl supermultiplets from the world volume of the instanton. The final result is~\cite{Beasley:2005iu}:
\be
\delta S = \int d^4 x  d^4 \theta {(W^2)}^g 
\exp{\left( -{{A(C)} \over {2 \pi \alpha'}} + i \int_C
B\right)}\,. \nonumber
\ee
where the sum over holomorphic curves $C$, accompanied by the exponential factor denoting the area of the
world sheet, defines the $A$-model amplitude $F_g$ at genus $g$, given in 
eqn. (\ref{Fg}). This is equivalent to the
computation in~\cite{9407190}, since in flat space-time, one needs graviton-graviphoton insertions in the
path integral to generate low energy F-terms. 

\subsection{$Z_{\rm BH}$}
$Z_{\rm BH}$ stands for the partition function of extremal black holes in four dimensions, which are solutions of $N=2$ supergravity coupled to $n_V$ vector multiplets with magnetic and electric charges $p^{\Lambda}, q_{\Lambda}$. 
The asymptotic values of the moduli fields 
$X^{\Lambda}$, $\Lambda = 0, 1,\cdots, n_V$ of the vector multiplets are arbitrary in the black hole solution. Near
the horizon of the black hole, the values of the moduli fields are constant and fixed by the 
attractor equations~\cite{Ferrara:1995ih}:
\be
p^{\Lambda} =  Re[CX^{\Lambda}], \qquad q_{\Lambda} = Re[CF_{\Lambda}]
\ee
where, $F_{\Lambda} = \frac{\partial F}{\partial X^{\Lambda}} $ and $C$ is a complex constant.
In the presence of higher derivative F-terms in 
the action, encoded in the prepotential:
\be
F(X^{\Lambda}, W^2) =  \sum_{g=0}^{\infty} F_g(X^{\Lambda}) W^{2g} \, ,
\ee
where $F_g$ are computed by eqn. (\ref{Fg}) and $W^2$ contains the square of the anti-self-dual
graviphoton field strength, the attractor equations are modified due to~\cite{Lopes Cardoso:1998wt,LopesCardoso:1999cv,Lopes Cardoso:1999ur}:
\be
C^2W^2 = 256 \, ,
\ee
in the gauge $K=0, C=2Q$. Here, $K$ denotes the K\"{a}hler potential and $Q$ is a complex combination of electric and
magnetic graviphoton charges. Thus, the black hole partition function takes the form~\cite{0405146}:
\be \label{zbh}
\ln Z_{BH} =  -4\pi Q^2 \left[\sum_g F_g \left(\frac{p^{\Lambda} + i \frac{\phi^{\Lambda}}{\pi}}{2Q} \right)\left(\frac{8}{Q}\right)^{2g}  \right]\, .
\ee
where, one uses $X^{\Lambda} = \frac{p^{\Lambda} + i \frac{\phi^{\Lambda}}{\pi}}{2Q}$. Here, $\phi^{\Lambda}$ are continuous electric potentials,
which are conjugate to integer electric charges $q^{\Lambda}$. As argued in~\cite{0608021,Guica:2007wd}, the partition function 
in eqn. (\ref{zbh}) is related to $Z_{\rm IIA}$ as in 
eqn. (\ref{zIIazbh}). The computation of $Z_{\rm IIA}$  involves taking in to account the
contributions of world-sheet instantons which wrap non-trivial curves
in the Calabi-Yau three-fold~\cite{Guica:2007wd,0608021,Beasley:2005iu}. 
\noindent


\section{Type IIA Partition Function on $AdS_2 \times S^2 \times {\rm CY}_3$} \label{ZIIA}

In this section, we calculate the partition function of type IIA superstrings on $AdS_2 \times S^2 \times {\rm CY}_3$ and show its connection to the $N=2$ topological string partition function. There are at least two methods to compute the partition function of type IIA superstrings on $AdS_2 \times S^2 \times {\rm CY}_3$. In the first method, using the fact that $c=6$ is the critical central charge for an $N=2$ matter system given in eqn. (\ref{ads2s2N2}), one introduces a set of $c= -6,\, N=2$ ghosts, constructs an $N=2$ BRST operator and 
calculates the partition function using standard $N=2$ techniques~\cite{Berkovits:1992bm}. 
The second method, is to twist the original $c= 6,\, N=2$ algebra of eqn. (\ref{ads2s2N2}), embed it in a small $N=4$ algebra and compute the partition function using the $N=4$ topological
techniques. This is the $N=4$ topological method we follow and is based on the $N=4$ topological strings discussed in section-\ref{N4topstrings}. The $N=4$ topological method can be used to calculate the partition function of any $N=2$ system which has a critical central charge of $c=6$~\cite{9407190}.

\subsection{Embedding $AdS_2 \times S^2 \times {\rm CY}_3$ in $N=4$ Topological Strings } \label{embedd}

As mentioned above, to use the $N=4$ topological method in the present case, one has to 
embedd the $AdS_2 \times S^2 \times {\rm CY}_3$ generators in
the twisted $N=4$ algebra. 

We start by explicitly denoting the $c=6,\, N=2$ algebra generators given 
in eqn. (\ref{ads2s2N2}) as: $T^L,G^{-L}, G^{+L}, J^L $ and 
$T^R,G^{-R}, G^{+R}, J^R$, where all the fermionic generators $G^{L,R\pm}$ have spin $\frac{3}{2}$. Now, there are two possible embeddings of the $AdS_2 \times S^2 \times {\rm CY}_3$
algebra in the $N=4$ topological algebra. In one case, we end up with an A-model in the 
Calabi-Yau discussed in section-\ref{N2topstrings}, and in the other case, the conjugate
${\bar A}$-model. We now discuss the A-model embedding. This can be achieved by the 
shifts, $T^L_{\rm new} = T^L - \frac{1}{2}\partial J^L$ and $T^R_{\rm new} = T^R - \frac{1}{2}\partial J^R$. This results in the
shift of spins $h^L,h^R$ of the generators of left and right-moving algebras as, $h^L_{\rm new} = h^L - \frac{1}{2}q^L$ and $h^R_{\rm new} = h^R - \frac{1}{2}q^R$, where $q^L,q^R$ 
are the $U(1)$ charges, respectively. As a result, the generators which acquire spin 1 and
spin 2 in the full $AdS_2 \times S^2 \times {\rm CY}_3$ geometry are, $G^{+L}, G^{+R}$ and $G^{-L}, G^{-R}$, respectively. Using eqns. (\ref{N2}), (\ref{N2r}) and (\ref{ads2s2N2}), this translates to $G^{+L}_{CY}, G^{-R}_{CY}$ acquiring spin 1
and $G^{-L}_{CY}, G^{+R}_{CY}$ acquiring spin 2 in the Calabi-Yau theory. This matches exactly with the A-twist discussed in section-\ref{N2topstrings}. Thus, we have an A-model topological
string in the Calabi-Yau geometry. On the other hand, in the $AdS_2 \times S^2$ geometry, it is $G^{+L}_{{\rm d}=4}, G^{+R}_{{\rm d}=4}$ that acquire spin 1 and $G^{-L}_{{\rm d}=4}, G^{-R}_{{\rm d}=4}$ acquire spin 2. We will have more to stay on this when we discuss the 
construction of BRST operator in section-\ref{localization}. To summarize, the twisting
of the algebra in the $AdS_2 \times S^2$ is opposite to that in the Calabi-Yau.

Now again the $T_{\rm new}(y)T_{\rm new}(z)$ OPE's have no central charge in the twisted $N=2$ system, and all bosonic and fermionic world sheet fields have integer spin. However, $G^+(y)G^-(z)$ OPE's have a central charge given as $\hat c= 2$. Since, $\hat c= 2$ is the critical central charge of an $N=4$ topological
algebra discussed in section-\ref{N4topstrings}, one can enlarge this twisted $N=2$ algebra, with two additional
currents of charge $\pm 2$ denoted by $J^{++}$ and $J^{--}$,
which together with $J$ satisfy an $SU(2)_c$ algebra, where the subscript $c$ stands for color.
Under
this $SU(2)_c$, $G^-$ and $G^+$ generate two new supercurrents $\tilde G^+$ and $\tilde G^-$.
Using the definitions in eqn. (\ref{ads2s2N2}) and the formulas in eqn. (\ref{Gtilde}), we can explicitly write:
\bea \label{Gtildeads}
\tilde G^- &=& e^{-2\rho-J_{CY}} ( {\bar d})^2 + e^{-\rho}\tilde G^{--}_{CY} , \cr
\tilde G^+ &=& e^{2\rho + J_{CY}} (d)^2 + e^{\rho}
\tilde G^{++}_{CY} \, ,
\eea
where, $\tilde G^{--}_{CY}$ and $\tilde G^{++}_{CY}$ are residues of the pole in the OPE of
$e^{J_{CY}}$ with $G^-_{CY}$ and $e^{-J_{CY}}$ with $G^+_{CY}$,
respectively. Their explicit form will not be needed here, but can be constructed using the methods in~\cite{Linch:2006ig}. The full set of generators for this $N=4$ system is already given in eqn. (\ref{N4fullgen}).

\subsection{$Z_{\rm IIA}$}

In the last subsection, we showed how the hybrid formalism in $AdS_2 \times S^2 \times {\rm CY}_3$ is a critical 
$N=2$ string theory and can be embedded in the $N=4$ topological algebra. This embedding in particular implies that, the partition function of IIA superstrings on $AdS_2 \times S^2 \times {\rm CY}_3$ can be calculated 
using eqn. (\ref{t4}).  

\noindent
One of the consequences of topological twisting is the existence of a nilpotent generator
$Q$. In terms of the twisting procedure discussed in last subsection, this generator can
be constructed from $G^{+L}$ and $G^{+R}$. The explicit form of this 
generator will not be needed here. However, the $AdS_2 \times S^2$ part of this generator
is important for localization arguments and will be discussed in section-\ref{localization}.
Due to the twisting, the type IIA path integral $Z_{\rm IIA}$, reduces to  a sum of local contributions from the fixed points of the nilpotent generator $Q$~\cite{Witten:1991zz}. In the Calabi-Yau geometry, these contributions come from configurations which are world sheet instantons in the A-model, and anti-instantons in the
conjugate ${\bar A}$-model. It is well known that the world-sheet instantons wrap non trivial curves in the Calabi-Yau and lead to holomorphic maps~\cite{Dine:1986zy}, given as:
\be
{\bar \partial} Y^i =0.
\ee
The anti-instantons come from anti-holomorphic maps in the Calabi-Yau. In $AdS_2 \times S^2$ geometry, such world sheet instanton configurations break translational symmetries and associated supersymmetries. After a euclidean rotation in the world sheet action in eqn. (\ref{actiond}), the instantonic configuration can be obtained by setting the world sheet variables $X^m$, $\theta^\a ,\theta^\ad$ and their right-moving counter parts to zero\footnote{For the four dimensional part, we sometimes use the flat-space notation of section-2.1, as the left-invariant currents of $AdS_2 \times S^2$ can always be written in terms of them, using the expansions in eqns. (\ref{Jone}) and (\ref{Jtwo}). }.  The only non-trivial terms
remaining in the $AdS_2 \times S^2$ action (\ref{actiond}) are:
\be \la{instg}
 \int  \left( d_{\alpha_L}d_{\beta_R} W^{\alpha_L\beta_R}  +   {\bar d}_{\dot\alpha_L} {\bar d}_{\dot\beta_R} W^{\dot\alpha_L\dot\beta_R}  \right)\, .
\ee
To summarize, $Z_{\rm IIA}$ reduces to a sum of 
two contributions, coming from world sheet instantons and anti-instantons\footnote{Let us note that, to reproduce the F-terms in eqn. (\ref{F}), one has to insert $2g-2$ chiral
graviphoton and two graviton vertex operators in the amplitude in flat space. In $AdS_2 \times S^2$, the relevant insertions are provided by the action of the world-sheet instantons in eqn. (\ref{instg}). 
}, which wrap holomorphic and anti-holomorphic curves in the Calabi-Yau, respectively and have legs in 
$AdS_2 \times S^2$. The action in eqn. (\ref{instg}) will be useful in the evaluation
of these contributions to the partition function. 

We now discuss the instanton contribution to $Z_{\rm IIA}$ and 
denote it as $Z^{\rm I}_{\rm IIA}$.
Towards the end,
we briefly discuss the anti-instanton contribution to the type IIA partition function, denoted as $Z^{\rm AI}_{\rm IIA}$. This corresponds to an opposite twist to the one in section-\ref{embedd}. The final answer for $Z_{\rm IIA}$ will be obtained by joining $Z^{\rm I}_{\rm IIA}$ and $Z^{\rm AI}_{\rm IIA}$. 

\begin{flushleft}
{\underline{Worldsheet Instantons}}
\end{flushleft}
With the choice of twisting in section-\ref{embedd}, eqn. (\ref{instantonP}) below corresponds to the instanton contribution to the type IIA partition function, $Z^{\rm I}_{\rm IIA}$. One
way to see this is that in the Calabi-Yau, we have an A-model topological string. The spin $2$ generators $G^{-}, \tilde G^{-}$ are like the $b$-ghost of the bosonic strings and
appear in the partition function in eqn. (\ref{instantonP}), integrated against the $3g-3$ Beltrami differentials. 
Thus using eqn. (\ref{t4}), the instanton contribution to the partition function of type IIA  superstrings on $AdS_2 \times S^2 \times {\rm CY}_3$ is:
\bea \label{instantonP}
&&Z^{\rm I}_{\rm IIA} = |\int du \sum_{n_I=2-2g}^{2g-2}
(u^*_2)^{2g-2+n_I} (u^*_1)^{2g-2-n_I}|^2 \cr
&& \prod_{j=1}^{3g-3}\int d^2 m_j
\prod_{i=1}^g \int d^2 v_i
<|\prod_{i=1}^{g-1} \widehat{{\tilde G}^+}(v_i) J(v_g)
(\int\mu_j \widehat{G^-})|^2> 
\eea
where, the $\widehat{{\tilde G}^+}$  and $\widehat{G^-}$
are defined in eqn. (\ref{su2f}). Just as in flat space~\cite{9407190}, the integration over $u$ can be done using,
\be
\int du  (u_1^{2g-2+m} u_2^{2g-2-n})
(u^*_2)^{2g-2+n} (u^*_1)^{2g-2-n} =
\delta_{mn}(2g-2+m)!
(2g-2-n)! \, .
\ee
To calculate $Z^{\rm I}_{\rm IIA}$ in hybrid formalism, the world sheet fields which we need to integrate over in the path integral are, $(a)$ $AdS_2 \times S^2$ fields: $X^m,(\theta^\a ,
d_{\a}), (\bar\theta^\ad, \bar d_{\ad})$  and their right-moving counter parts, for $m=0,\cdots,3$ and $\a,\ad = 1,2$, $(b)$ Calabi-Yau variables and, $(c)$ the chiral boson $\rho$ and its right-moving counter part. 
 
Let us start by choosing $n_I = g-1$ in eqn. (\ref{instantonP}), which was argued in~\cite{9407190}, to be the piece relevant for computing the 
low energy F-terms in eqn. (\ref{F}). Next, it is needed to
write the partition function (\ref{instantonP}), transforming variables from the hatted generators, 
$\widehat{{\tilde G}^+}$ and $\widehat{G^-}$, using the definitions in eqn. (\ref{su2f}). This 
can be done as follows. There are $3g-3$ $~\widehat{G^-}$'s 
and $g-1$ $~\widehat{{\tilde G}^+}$'s in the partition function
(\ref{instantonP}). Thus, using (\ref{su2f}), $\widehat{G^-}$'s can contribute
$~3g-3-l$ $~G^-$'s and $+l~$ $~{\tilde G}^-$'s, where as, 
$\widehat{{\tilde G}^+}$'s can contribute $g-1-l~$ $~G^+$'s and $+l~$ $~{\tilde G}^+$'s, in the partition function. These 
contributions are further subject to the constraints coming from
the background charges of the $U(1)$ current $J$. For the partition function to not vanish trivially, one has to ensure
that $~G^{\pm}$ 
and $~{\tilde G}^{\pm}$ contributions contain $1-g$ units
of $\rho$ charge and $3g-3$ units of $J_{CY}$ charge~\cite{9407190}. 
This is only possible if, each $G^-$ contributes $G_{CY}^-$, each $\tilde G^-$
contributes
$e^{-2\rho-\int J_{CY}} (\bar d)^2$, each $G^+$ contributes $(\bar d)^2 e^{-\rho}$,
and each $\tilde G^+$ contributes $\tilde G_{CY}^{++} e^{\rho}$.
Making these choices, we have\footnote{We omit the details
of the regularization over the negative energy of the chiral boson, which is same as in flat space and discussed in detail in~\cite{9407190}. The regularization fixes the
the poles of $\,e^{\rho} {\tilde G}^{++}_{\rm CY}$ coming from ${\tilde G}^+$ to be sewn in with those of
$ e^{-2\rho - \int J_{\rm CY}}({\bar d})^2$ coming from ${\tilde G}^-$, giving the term $\prod_{j=1}^l   (~e^{-\rho}({\bar d})^2 \int\mu_j  G^{-}_{\rm CY} ~)$ in eqn. (\ref{instantonP2}). It also contributes a factor of
$\det({\rm Im}\tau) |\det \omega^k(v_i)|^2 $, from the Jacobian of change of variables, allowing the
$v_i$'s to be any $g$ points on $\Sigma_g$.}:
\bea \label{instantonP2}
&&Z^{\rm I}_{\rm IIA} =
\prod_{j=1}^{3g-3}\int d^2 m_j  \det({\rm Im}\tau) |\det \omega^k(v_i)|^2 \cr
&&\left< \left|\prod_{i=l+1}^{g-1} e^{-\rho}({\bar d})^2(v_i) J(v_g)~
\prod_{j=1}^l   (~e^{-\rho}({\bar d})^2 \int\mu_j  G^{-}_{\rm CY} ~)~
\prod_{k=l+1}^{3g-3}(\int\mu_k G^-_{\rm CY}~)(.1)\right|^2 \right>.
\eea
Here, $\tau$ is the period matrix and $\omega^k$ are $g$ holomorphic one forms on $\Sigma_g$. We have also
inserted $1 = \int (G^+ , {\bar\theta}^2 e^{\rho}) $ in eqn. (\ref{instantonP2}). This makes the functional integral well defined and can be used to remove the factor of $J(v_g)$, as follows.
Using the definitions in eqn. (\ref{ads2s2N2}), notice that $G^+$ has OPE's only with 
$J(v_g)$ in eqn. (\ref{instantonP2}). Thus, one can pull the
$\int G^+$ counter off the
term ${\bar\theta}^2 e^{\rho}$, till it hits $J(v_g)$, giving:
\bea \label{instantonP3}
&&Z^{\rm I}_{\rm IIA} =
\prod_{j=1}^{3g-3}\int d^2 m_j  \det({\rm Im}\tau) |\det \omega^k(v_i)|^2 \cr
&&\left< \left|\prod_{i=l+1}^{g} e^{-\rho}({\bar d})^2(v_i)~
\prod_{j=1}^l   (~e^{-\rho}({\bar d})^2 \int\mu_j  G^{-}_{\rm CY} ~)~
\prod_{k=l+1}^{3g-3}(\int\mu_k G^-_{\rm CY})~({\bar\theta}^2 e^{\rho})\right|^2 \right>.
\eea

\noindent
In eqn. (\ref{instantonP3}), apart from functional integral over
various world sheet coordinates, one also has to perform integrations
over the zero modes of these variables. The counting of zero modes is as follows. There are four bosonic zero modes of
$X^m$ and eight fermionic zero modes corresponding to $\theta,\bar \theta$'s. Further, on a genus $g$ Riemann surface,
each variable $d$ contributes $g$-zero modes, giving in total,
$8g$-zero modes. These zero modes are saturated as follows. In eqn. (\ref{instantonP3}), four zero modes of
$\bar \theta$'s are explicitly present together with $4g$-zero
of $\bar d$'s. Another $4g$-zero modes of $d$'s can also be pulled
out from $2g$ powers of the first term in the instanton action in eqn. (\ref{instg}) as:
\be \la{instgloops}
\int  \left(d_{\alpha_L}d_{\beta_R} \delta^{\alpha_L\beta_R} \,N\, g_s\,\right)^{2g} \, .
\ee
where, in an extremal black hole background, the expectation value of the
graviphoton field strength is
$W^{\alpha_L\beta_R} = \delta^{\alpha_L\beta_R} \,N\, g_s\, $.
These exactly saturate the required zero modes of $d$'s, in the partition function in eqn. (\ref{instantonP3}). Thus we have:
\bea \label{instantonP4}
Z^{\rm I}_{\rm IIA} &=&
\prod_{j=1}^{3g-3}\int d^2 m_j  \det({\rm Im}\tau) |\det \omega^k(v_i)|^2 \cr
&\times &\left< \left|\prod_{i=l+1}^{g} e^{-\rho}({\bar d})^2(v_i)~
\prod_{j=1}^l   (~e^{-\rho}({\bar d})^2 \int\mu_j  G^{-}_{\rm CY} ~)~
\prod_{k=l+1}^{3g-3}(\int\mu_k G^-_{\rm CY})~({\bar\theta}^2 e^{\rho})\right|^2 \right. \cr
&\times& \left. (g_{top})^{2g} \, (\prod_{r=1}^{2g}\int d^2 z_r d^2)  \right> , \\
&=& 
 \prod_{j=1}^{3g-3}\int d^2 m_j  \det({\rm Im}\tau) |\det \omega^k(v_i)|^2 \cr
&\times & \left< \left|\prod_{i=l+1}^{g} e^{-\rho}({\bar d})^2(v_i)~
\prod_{j=1}^l   (~e^{-\rho}({\bar d})^2) 
~({\bar\theta}^2 e^{\rho}) \right|^2
(g_{top})^{2g} \, (\prod_{r=1}^{2g}\int d^2 z_r d^2) \right> \cr
&\times & \left<\left| (\prod_{j=1}^{3g-3} \int\mu_j  G^{-}_{\rm CY} ~)~  \right|^2\right> \label{instantonP5}.
\eea
where $g_{top} = N\, g_s$, and in the last line of eqn. (\ref{instantonP5}) the Calabi-Yau part is separated. 

In eqn. (\ref{instantonP5}), the integrations over 
various worlds-sheet coordinates can now be done exactly as in
flat space time~\cite{9407190}. The functional integrals over the $\rho$ and
$(\bar d_1, \bar \theta^{1})$ fields
contribute $|Z_1|^{-2}[det(Im\tau)]^{-1}$,
$(\bar p_2,\bar\theta^{2})$ fields contributes $|Z_1 det \omega_j(v_k)|^2$, $(d_\a,\theta^\a)$ contribute $|Z_1|^4
(det[Im\tau])^2$ and $x^m$'s give a factor
of \\$|Z_1|^{-4}(det[Im\tau])^{-2}$, where,
$(Z_1)^{-\half}$ is the partition function 
for a chiral boson~\cite{Verlinde:1987sd}. Thus, we have:
\be \label{zeroM}
Z^{\rm I}_{\rm IIA} = g_{\rm top}^{2g}
\int d^4X d^2\theta_L  d^2\theta_R 
\prod_{j=1}^{3g-3}\int d^2 m_j \left<\left| (\prod_{j=1}^{3g-3} \int\mu_j  G^{-}_{\rm CY} ~)~  \right|^2\right> ,
\ee
where, explicitly, the A-model topological string amplitude as given in eqn. (\ref{Fg}) is:
\be \la{topA}
F_g = \prod_{j=1}^{3g-3}\int d^2 m_j\left< \prod_{j=1}^{3g-3} (\int\mu^L_j G^{-L}_{\rm CY})(\int\mu^R_j G^{+R}_{\rm CY}) \right> \, .
\ee
Let us emphasize that, out of the eight fermionic 
zero modes required for a non zero result, the $|\bar \theta|^2$
term in eqn. (\ref{instantonP5}), has been used
to absorb the $(\bar\theta_{L\dot\alpha}, \bar\theta_{R\dot\alpha})$ integrals. One is left with
the integrations over four bosonic and fermionic variables in (\ref{zeroM}) as there is no action for these variables. In the following subsection, localization arguments are used to do
the integrations over these variables. 

\begin{flushleft}
{\underline{Anti-instantons}}
\end{flushleft}
A similar analysis as above can be repeated for the contribution of anti-instantons, which results in an ${\rm {\bar A}}$-topological model in the Calabi-Yau. In this
case, the anti-instanton contribution to the IIA partition function 
with ${\rm {\bar A}}$-twist is:
\bea \label{instantonantiP}
&&Z^{\rm AI}_{\rm IIA} = |\int du \sum_{n_I=2-2g}^{2g-2}
(u^*_2)^{2g-2+n_I} (u^*_1)^{2g-2-n_I}|^2 \cr
&& \prod_{j=1}^{3g-3}\int d^2 m_j
\prod_{i=1}^g \int d^2 v_i
<|\prod_{i=1}^{g-1} \widehat{{\tilde G}^-}(v_i) J(v_g)
(\int\mu_j \widehat{G^+})|^2> 
\eea
where, the $\widehat{{\tilde G}^-}$  and $\widehat{G^+}$
are defined in eqn. (\ref{su2f})
In this case, to do the integration over the $(\theta_{L\alpha},\theta_{R\alpha})$ zero modes,
$1 = \int (G^- ,{\theta}^2 e^{-\rho})$ is inserted in the path integral and the counter pulling argument over $J(v_g)$ is
repeated. The absorption of 4g-zero modes of $\bar d$'s requires pulling 2g powers of the
second term in the instanton action in eqn. (\ref{instg}). Finally, one is left with the integration over four $X_m$ and $(\bar\theta_{L\dot\alpha}, \bar\theta_{R\dot\alpha})$ zero modes as:
\be \label{zeroAM}
{\bar Z}^{\rm AI}_{\rm IIA} = g_{\rm top}^{2g}
\int d^4X d^2\bar\theta_L  d^2\bar\theta_R 
\prod_{j=1}^{3g-3}\int d^2 m_j \left<\left| (\prod_{j=1}^{3g-3} \int\mu_j  G^{+}_{\rm CY} ~)~  \right|^2\right> ,
\ee
where the ${\rm \bar A}$-model topological string amplitude is:
\be \la{topAbar}
\bar F_g = \prod_{j=1}^{3g-3}\int d^2 m_j \left< \prod_{j=1}^{3g-3} (\int\mu^L_j G^{+L}_{\rm CY})(\int\mu^R_j G^{-R}_{\rm CY}) \right> \, .
\ee
For this case, the opposite twisting leads to $G^{+L}_{\rm CY}$
and $G^{-R}_{\rm CY}$ acquiring spin $2$. A nilpotent operator in 
this case in the Calabi-Yau is:
\be \la{brstcy}
{\bar Q}_{\rm CY} = G^{-L}_{\rm CY} + G^{+R}_{\rm CY}.
\ee
The full partition function of type IIA superstrings on 
 $AdS_2 \times S^2 \times {\rm CY}_3$ is obtained from eqns. (\ref{zeroM}) and (\ref{zeroAM}). 

\subsection{BRST Localization} \label{localization}

In this section, our aim is perform the integration over $X_m$, $(\theta_{L\alpha},\theta_{R\alpha})$ and $(\bar\theta_{L\dot\alpha}, \bar\theta_{R\dot\alpha})$, appearing in eqns. (\ref{zeroM}) and (\ref{zeroAM}). In the present situation, the world sheet instanton in $AdS_2 \times S^2$ contributes four bosonic 
zero modes coming from the breaking of translation isometries and their supersymmetric 
partners. From the $PSU(1,1|2)$ sigma model point of view, the left-multiplication symmetry
corresponding to the generators $T_m$ and $T_{\a,\ad}$ is broken by the instanton expectation
value.
Since, there is no action for $X_m$ in eqns. (\ref{zeroM}) and (\ref{zeroAM}), the integration
gives rise to an infinite space-time volume of $AdS_2$. Also, since there is no action for $(\theta_{L\alpha},\theta_{R\alpha})$ and $(\bar\theta_{L\dot\alpha}, \bar\theta_{R\dot\alpha})$, the result is zero. Below, we use a localization procedure similar to~\cite{0608021} to solve this problem.

Let us note that, in the Green-Schwarz treatment~\cite{0608021}, due to the
presence of $\kappa$-symmetry, a world-sheet instanton preserves four out of eight 
space-time supersymmetries. In particular, instantons and anti-instantons at the opposite
poles of $S^2$ preserve same amount of supersymmetry, but a different set, namely chiral and 
anti-chiral, respectively. This reflects in our analysis in eqns. (\ref{zeroM}) and (\ref{zeroAM}). In eqn. (\ref{zeroM}),  one is left with integration over chiral set of variables $(\theta_{L\alpha},\theta_{R\alpha})$ of the instantons and in eqn. (\ref{zeroAM}),
over the anti-chiral variables corresponding to anti-instantons.
In the 
hybrid description of the superstring such (anti)instantons break all space-time
supersymmetries, since one cannot perform a compensating $\kappa$-transformation to be in
the gauge. Instead, one has superconformal invariance, which can be used to write down a 
operator which renders the partition function finite. The 
advantage of the hybrid description is that a unified BRST treatment can be given for $AdS_2 \times S^2$ and the 
Calabi-Yau parts, using the nilpotent operator $Q$ discussed at the beginning 
of section-\ref{ZIIA} . Since, the $AdS_2 \times S^2$ and the Calabi-Yau CFT's decouple, 
$Q$ contains four and six dimensional parts. Below, we explicitly give the 
$AdS_2 \times S^2$ part of this operator denoted as $Q_{{\rm d} = 4}$ for localizing world-sheet instantons and ${\bar Q}_{{\rm d} = 4}$ for anti-instantons. These operators turn out to be
BRST operators of the $AdS_2 \times S^2$ sigma model.

In the present case, the essence of the localization procedure is to use these BRST operators to make a semi-classical expansion of the path integral around the saddle points. 
This is possible, because the $PSU(1,1|2)$ sigma model action
can be shown to be exact under these BRST operators, constructed from the left-invariant currents. Thus, if the action $S_{AdS_2 \times S^2}$ is ${\bar Q}_{{\rm d} = 4}$-closed and 
${\bar Q}_{{\rm d} = 4}^2 = 0$, then one can deform the action by adding another
BRST exact term as:
\be
S_{AdS_2 \times S^2} =  S_{AdS_2 \times S^2} + t\, {\bar I}
\ee
where ${\bar I} = {\bar Q}_{{\rm d} = 4} {\bar V}$ is an arbitrary parameter for some choice of $ {\bar V}$. Adding such a term does not 
change the cohomology and the parameter $t$ can be varied freely~\cite{Witten:1988xj,Pestun:2007rz}. 
In particular, as $t \rightarrow \infty$, the
theory localizes to the critical points of ${\bar Q}_{{\rm d} = 4}{\bar V}$. Thus, one can integrate over these critical points to
render the path integral in eqn. (\ref{zeroAM}) finite, by localizing the 
genus $g$ anti-instantons at the (north) of $S^2$. The analysis can be repeated using
$Q_{{\rm d} = 4}$  for integration over variables in eqn. (\ref{zeroAM}).

We now
write ${\bar Q}_{{\rm d} = 4}$ and
then show that the sigma model action is
${\bar Q}_{{\rm d} = 4}$-closed. For this purpose, it is useful to use
the $U(1) \times U(1)$ form of the $AdS_2 \times S^2$ action
derived in section-\ref{u1notation} and write\footnote{The choice of this operator coincides with the fact that the instanton and the anti-instanton require
integration over fermionic variables with different set of indices, namely, undotted and dotted. This is also related to
the fact that in euclidean $AdS_2 \times S^2$, the Symplectic Majorana condition on fermions, does not take dotted in to undotted indices. For instance, $(\theta^{\alpha}_{i})^{\dagger} = \epsilon^{\alpha\beta}\epsilon_{ij}\theta_{\beta}^j$ and
$(\theta^{\ad}_{i})^{\dagger} = \epsilon^{\ad\bd}\epsilon_{ij}\theta_{\bd}^j$, where $\a,\ad,\b,\bd = 1,2$ and $i,j = L,R$ (see~\cite{0608021} for related discussion). }:
\be \la{brst}
{\bar Q}_{{\rm d} = 4} =  G^{-L}_{{\rm d} = 4} + G^{-R}_{{\rm d =4}}.
\ee
In terms of left-invariant currents, we have
\bea
G^{-L}_{{\rm d} = 4} &=&  \int \, e^{\rho_L} \,{\bar J}_L^{++} {\bar J}_L^{--}  \, , \cr
G^{-R}_{{\rm d =4}} &=&  \int \, e^{\rho_R} \,J_R^{++}J_R^{--}  \, .
\eea
The operator ${\bar Q}_{{\rm d} = 4}$ in eqn. (\ref{brst}) is nilpotent, i.e.,
${\bar Q}_{{\rm d} = 4}^2 = 0$ as it is constructed from a sum of $G^{-L}_{{\rm d} = 4}$ and $G^{-R}_{{\rm d =4}}$, which have regular
OPE's of the $N=2$ algebra. Also, using the equations of motion in (\ref{eomads2s2}), one can check that $G^{-L}_{{\rm d} = 4}$
is anti-holomorphic and $G^{-R}_{{\rm d =4}}$ is holomorphic.
The BRST operator is constructed from covariantly holomorphic and anti-holomorphic H-invariant combination of currents~\cite{9907200}. The action of ${\bar Q}_{{\rm d} = 4}$ on the group element ${\rm g}$ 
is to transform it by a right multiplication as:
\be \label{Qg}
\Lambda {\bar Q}_{{\rm d} = 4} ({\rm g}) =   \,{\rm g} \left( \Lambda\, (e^{\rho}\, {\bar J}_3) + \Lambda \, (e^{\rho}\, J_1) \right) \, ,
\ee
where $\Lambda$ is a constant anti-commuting parameter. 
The left-invariant currents transform under (\ref{Qg}) as
\bea
{\bar Q}_{{\rm d} = 4} (J_j) &=& \d_{j+1,0} ~\partial(\Lambda\, e^{\rho}\, {\bar J}_3) + [J_{j+1},\Lambda\, e^{\rho}\, {\bar J}_3] \cr 
&~& ~~~~~~~~~~~~~~~~~~~~~+\d_{j+3,0}~\partial(\Lambda \, e^{\rho}\, J_1 ) + 
[J_{j+3},\Lambda \, e^{\rho}\, J_1 ] \, , \cr
{\bar Q}_{{\rm d} = 4} (\bar J_j ) &=& \d_{j+1,0}~\bar\partial(\Lambda\, e^{\rho}\, {\bar J}_3) + [\bar J_{j+1},\Lambda\, e^{\rho}\, {\bar J}_3] 
\cr
&~& ~~~~~~~~~~~~~~~~~~~~~+\d_{j+3,0}~\bar\partial(\Lambda \, e^{\rho}\, J_1 ) +
 [\bar J_{j+3},\Lambda \, e^{\rho}\, J_1 ], 
\eea
where $j$ is defined modulo 4, i.e. $J_j \equiv J_{j+4}$. 
Now, one can show that the $PSU(1,1|2)$ sigma model action in (\ref{cosetU}) is closed under the BRST 
operator (\ref{brst}). To check this, we follow the procedure used in~\cite{Berkovits:2004xu}, and start by rewriting the sigma model action in eqn. (\ref{cosetU}) as:
\bea \label{cosetbrst}
S_{AdS_2 \times S^2} &=& 
\int d^2z \left[ \; \frac{1}{2} \left( J^{\pp 0}{\bar J}^{\mm 0} - J^{0\pp}{\bar J}^{0\mm} 
+ J^{\mm 0}{\bar J}^{\pp 0} - J^{0\mm}{\bar J}^{0\pp} \right) + \frac{1}{2} \left( - J_L^{--}{\bar J}_R^{++}
 \right. \right. \cr
&-& \left. \left.
{\bar J}_L^{--} J_R^{++}  - J_L^{++}{\bar J}_R^{--}
- {\bar J}_L^{++} J_R^{--}  \right) + \frac{1}{2} \left( J_L^{+-}{\bar J}_R^{-+} + {\bar J}_L^{+-} J_R^{-+} 
+ J_L^{-+}{\bar J}_R^{+-}\right. \right. \cr
&+& \left. \left.
{\bar J}_L^{-+} J_R^{+-} \right) + \frac{1}{4}
\left( J_L^{--}{\bar J}_R^{++}
- {\bar J}_L^{--} J_R^{++} + J_L^{++}{\bar J}_R^{--}
- {\bar J}_L^{++} J_R^{--} \right) \right. \cr
&+& \left. \frac{1}{4}
\left( -J_L^{+-}{\bar J}_R^{-+}
+ {\bar J}_L^{+-} J_R^{-+} - J_L^{-+}{\bar J}_R^{+-}
+  {\bar J}_L^{-+} J_R^{+-}\right)
\right]
\eea
Now, under the BRST operator (\ref{brst}), the variations of the terms in first and the third brackets in (\ref{cosetbrst}) can be
shown to cancel each other. The variation of the terms in second bracket is,
\bea \label{second}
&& <\,-\frac{1}{2}\left[ 
\nabla(\Lambda e^{\rho_L} {\bar J}_L^{--}) {\bar J}_R^{++} + J_L^{--} {\bar \nabla}(\Lambda e^{\rho_R} J_R^{++})
+{\bar\nabla}(\Lambda e^{\rho_L} {\bar J}_L^{--})  J_R^{++}  \right. \cr
&& \left. + {\bar J}_L^{--} \nabla(\Lambda e^{\rho_R} J_R^{++}) +\nabla(\Lambda e^{\rho_L} {\bar J}_L^{++}) {\bar J}_R^{--} + J_L^{++} {\bar \nabla}(\Lambda e^{\rho_R} J_R^{--}) \right. \cr
&& \left. +{\bar\nabla}(\Lambda e^{\rho_L} {\bar J}_L^{++})  J_R^{--} + {\bar J}_L^{++} \nabla(\Lambda e^{\rho_R} J_R^{--})
\right]\,> \, ,
\eea
and the variations of the terms in the fourth and fifth brackets are respectively:
\bea \label{four}
&&<\,\frac{1}{4}\left[ 
\nabla(\Lambda e^{\rho_L} {\bar J}_L^{--}) {\bar J}_R^{++} + J_L^{--} {\bar \nabla}(\Lambda e^{\rho_R} J_R^{++})
-{\bar\nabla}(\Lambda e^{\rho_L} {\bar J}_L^{--})  J_R^{++}  \right. \cr
&& \left. - {\bar J}_L^{--} \nabla(\Lambda e^{\rho_R} J_R^{++}) +\nabla(\Lambda e^{\rho_L} {\bar J}_L^{++}) {\bar J}_R^{--} + J_L^{++} {\bar \nabla}(\Lambda e^{\rho_R} J_R^{--}) \right. \cr
&& \left. -{\bar\nabla}(\Lambda e^{\rho_L} {\bar J}_L^{++})  J_R^{--} - {\bar J}_L^{++} \nabla(\Lambda e^{\rho_R} J_R^{--})
\right]\, > \, ,
\eea
and,
\bea \label{five}
&&<\,\frac{1}{4}\left[ 
-(J^{\pp 0}\Lambda e^{\rho_R} J_R^{--} - J^{0\mm}\Lambda e^{\rho_R} J_R^{++} ) {\bar J}_R^{-+} 
- J_L^{+-}(\bar J^{0\pp}\Lambda e^{\rho_L} \bar J_L^{--} - \bar J^{\mm 0}\Lambda e^{\rho_L} \bar J_L^{++} ) \right. \cr
&& \left. -(-J^{0\pp }\Lambda e^{\rho_R} J_R^{--} + J^{\mm 0}\Lambda e^{\rho_R} J_R^{++} ) {\bar J}_R^{+-} 
- J_L^{-+}(\bar J^{0\mm}\Lambda e^{\rho_L} \bar J_L^{++} - \bar J^{\pp 0}\Lambda e^{\rho_L} \bar J_L^{--} ) \right. \cr
&&\left. +( \bar J^{\pp 0}\Lambda e^{\rho_R} J_R^{--} - \bar J^{0\mm}\Lambda e^{\rho_R} J_R^{++} ) J_R^{-+} 
+ \bar J_L^{+-}( J^{0\pp}\Lambda e^{\rho_L} \bar J_L^{--} - J^{\mm 0}\Lambda e^{\rho_L} \bar J_L^{++} ) \right. \cr
&& \left. +(- \bar J^{0\pp }\Lambda e^{\rho_R} J_R^{--} + \bar J^{\mm 0}\Lambda e^{\rho_R} J_R^{++} )  J_R^{+-} \right. \cr
&& \left. +\bar J_L^{-+}( J^{0\mm}\Lambda e^{\rho_L} \bar J_L^{++} - J^{\pp 0}\Lambda e^{\rho_L} \bar J_L^{--} )
\right] \, >\, .
\eea
Now, one makes use of the Maurer-Cartan equations coming from $\partial \bar J - \bar\partial J + [J, \bar J] = 0$. 
For instance, we have:
\be
\nabla \bar J_L^{--} - \bar\nabla J_L^{--} + [J^{\mm 0}, \bar J_R^{+-}] + [ J_R^{+-}, \bar J^{\mm 0}] 
+ [J^{0\mm }, \bar J_R^{-+}] + [ J_R^{-+}, \bar J^{0\mm }]  = 0\,  .
\ee
Using 
these relations in (\ref{five}), one can show that the equations (\ref{four}) and (\ref{five}) are exactly equal.  
Thus, adding eqns. (\ref{second}) + (\ref{four}) + (\ref{five})
and doing some partial integrations, we get the full variation of (\ref{cosetbrst}) to be:
\be
<-
(\Lambda e^{\rho_L} {\bar J}_L^{--})\, {\bar\nabla} J_R^{++} + \nabla{\bar J}_L^{--} (\Lambda e^{\rho_R} J_R^{++})
 +(\Lambda e^{\rho_L} {\bar J}_L^{++})\,{\bar\nabla}  J_R^{--} + \nabla{\bar J}_L^{++} (\Lambda e^{\rho_R} J_R^{--})>\, .
\ee
This is identically zero, using the equations of motion in
(\ref{eomads2s2}). Thus we have shown that the $AdS_2 \times S^2$ sigma model action in (\ref{cosetU}), is closed under the BRST 
operator (\ref{brst}).\\

\noindent

Now, to do the $X_m$ and $(\bar\theta_{L\dot\alpha}, \bar\theta_{R\dot\alpha})$ integrations in (\ref{zeroAM}), a BRST exact operator is constructed as follows. 
We require an operator $I$, such that ${{\bar Q}_{{\rm d} = 4}\, \bar I} = 0$ and $\bar I = {{\bar Q}_{{\rm d} = 4} \, \bar V}$, which facilitates a simple Gaussian integration
over the variables. Below, it is shown that the general form of such an operator is given by:
\be \label{sample}
\bar V = (a (X_m)^2 + b \epsilon_{\dot\alpha\dot\beta}\bar\theta^{\dot\alpha}_L\bar\theta^{\dot\beta}_R) (e^{-\rho_L}\theta_R^2 + e^{-\rho_R} \theta_L^2)
\ee
where $m=0,1,2,3$, and the constants $a$ and $b$, are fixed by the BRST procedure. 

\be \label{triv}
\int e^{\rho_L} \,{\bar J}_L^{++}{\bar J}_L^{--} (e^{-\rho_L} \theta_R^2) = 1 \, ,\qquad
\int e^{\rho_R} \,J_R^{++}J_R^{--} (e^{-\rho_R} \theta_L^2) = 1 \, .
\ee
render the following choice of the operator,
\be \label{v}
\bar V = - \frac{1}{2}\left[( X^{\pp 0}X^{\mm 0} - X^{0\pp} X^{0\mm} + \theta_L^{+-}\theta_R^{-+} - \theta_L^{-+}\theta_R^{+-})\,(e^{-\rho_L} \theta_R^2 + e^{-\rho_R} \theta_L^2)\right] \, ,
\ee
a trivial deformation of the theory. To see this, one first notes that:
\bea \label{acting}
{\bar Q}_{{\rm d} = 4} \bar V &=& -\frac{1}{2}\left[ {\bar Q}  ( X^{\pp 0}X^{\mm 0} - X^{0\pp} X^{0\mm} + \theta_L^{+-}\theta_R^{-+} - \theta_L^{-+}\theta_R^{+-})\,\right] (e^{-\rho_L} \theta_R^2 + e^{-\rho_R} \theta_L^2) \cr
&-& \frac{1}{2}( X^{\pp 0}X^{\mm 0} - X^{0\pp} X^{0\mm} + \theta_L^{+-}\theta_R^{-+} - \theta_L^{-+}\theta_R^{+-})\,\left[ {\bar Q} (e^{-\rho_L} \theta_R^2 + e^{-\rho_R} \theta_L^2)\right] \, ,
\eea
Now, it can be shown that,
\be \label{actQ}
{\bar Q}_{{\rm d} = 4} ( X^{\pp 0}X^{\mm 0} - X^{0\pp} X^{0\mm} + \theta_L^{+-}\theta_R^{-+} - \theta_L^{-+}\theta_R^{+-}) =0 \, .
\ee
To check this explicitly, one can calculate how the left-invariant currents rotate various coordinates appearing in (\ref{v}). Thus, under $J_L^{--}$ and $J_L^{++}$, we have:
\bea
\delta\theta_L^{--} &=& \epsilon_L^{--} \, ,\cr
\delta \theta_L^{+-} &=&  \,\epsilon_L^{--} X^{\pp 0} , \qquad 
\delta \theta_L^{-+} =  -\,\epsilon_L^{--}X^{0\pp } \, , \cr
\delta X^{\mm 0} &=&  -\,\epsilon_L^{--}\theta_R^{-+}  , \qquad 
\eea
and, 
\bea
\delta\theta_L^{++} &=& \epsilon_L^{++}, \cr
\delta \theta_L^{+-} &=&  -\,\epsilon_L^{++} X^{0\mm} , \qquad 
\delta \theta_L^{-+} =  \,\epsilon_L^{++}X^{\mm 0}  \cr
\delta X^{\pp 0} &=&  -\,\epsilon_L^{++}\theta_R^{+-}  , \qquad 
\delta  X^{0\pp} =  \,\epsilon_L^{++} \theta_R^{-+} \, ,
\eea
and the transformations with respect to $\epsilon_R^{++},\epsilon_R^{--}$ can be obtained similarly. So, the
first term in (\ref{acting}) drops out. Using the relations in (\ref{triv}), one gets:
\be
\bar I = -( X^{\pp 0}X^{\mm 0} - X^{0\pp} X^{0\mm} + \theta_L^{+-}\theta_R^{-+} - \theta_L^{-+}\theta_R^{+-})
\ee
This is the required operator, as it also satisfies ${\bar Q} \, \bar I =0$, as already shown above. 

The BRST exact term for the case of instantons can be written down in a similar manner. In this case, we
need to do integrations over four $X_m$ and $(\theta_{L\alpha}, \theta_{R\alpha})$ zero modes in (\ref{zeroM}). 
\be \la{brstI}
Q_{{\rm d} = 4} =  G^{+L}_{{\rm d} = 4} + G^{+R}_{{\rm d =4}}.
\ee
In terms of left-invariant currents, we have,
\bea
G^{+L}_{{\rm d} = 4} &=&  \int \, e^{-\rho_L} \,{\bar J}_L^{+-} {\bar J}_L^{-+}  \, , \cr
G^{+R}_{{\rm d =4}} &=&  \int \, e^{-\rho_R} \,J_R^{+-}J_R^{-+}  \, .
\eea
The operator $Q_{{\rm d} = 4}$ in eqn. (\ref{brstI}) is also nilpotent, as it is constructed from a sum of $G^{+L}_{{\rm d} = 4}$ and $G^{+R}_{{\rm d =4}}$, which have regular
OPE's of the $N=2$ algebra.
The action can again be shown to be closed under $Q_{{\rm d} = 4}$ as shown for the case of anti-instantons. 
Thus, one now adds a term coming from,
\be \label{trivialAnti}
I =  -\left( X^{\pp 0}X^{\mm 0} - X^{0\pp} X^{0\mm} + \theta_L^{++}\theta_R^{--} - \theta_L^{--}\theta_R^{++}\right),
\ee
to the path integral in (\ref{zeroM}) as:  
\be \label{integral}
\int d^4X d^2\theta_L  d^2\theta_R e^{-t I} \, ,
\ee
Note that $I$ satisfies $Q_{{\rm d} = 4} \, I =0$ and $I = Q_{{\rm d} = 4} V$ for a suitable $V$. To proceed,
one  makes a euclidean rotation to write the bosonic part of $I$ as $ X^{\pp 0}X^{\mm 0} + X^{0\pp} X^{0\mm}$.
The euclidean $AdS_2 \times S^2$ can be taken as in~\cite{0608021} to be, 
\be
ds^2 = \frac{4 dw d{\bar w}}{(1-w{\bar w})^2 } + \frac{4 dz d\bar{z}}{(1+z\bar{z})^2} \, ,
\ee
and using the embedding,
\be
X^{\pp 0}X^{\mm 0} = \frac{|\omega|^2}{(1-|\omega|^2)^2} \, , \qquad 
X^{0\pp} X^{0\mm} = \frac{|z|^2}{(1+|z|^2)^2} \, ,
\ee
the integral in (\ref{integral}) is,
\be \label{int}
\int d^2\omega d^2z d\theta_L^{++}d\theta_L^{--}d\theta_R^{--}d\theta_R^{++} e^{-t \left(\frac{(|\omega|^2 + |z|^2((1 + |\omega|^2|z|^2)}{(1-|\omega|^2)^2(1+|z|^2)^2} 
- \theta_L^{++}\theta_R^{--} + \theta_L^{--}\theta_R^{++}\right)}
\ee
Now, since $Q_{{\rm d} = 4} \, I =0$, the expectation value of any $Q_{{\rm d} = 4}$-invariant operator is zero, as can be seen by differentiating
with respect to $t$. Thus, the parameter $t$ can be varied freely. 
Consider, $t \rightarrow \infty$ limit; the leading contributions to (\ref{int}) come from the center of $AdS_2$ and
the south pole of $S^2$, i.e., $|\omega|=0,|z| =0$. Thus, the bosonic integral becomes $\int d^2\omega d^2z e^{-t (|\omega|^2 + |z|^2)} $, giving the regularized value $\sim \pi^2/t^2$. This then also localizes the fermionic path integral to
these locations as well, giving the regularized value:
\be
\sim \int  d\theta_L^{++}d\theta_L^{--}d\theta_R^{--}d\theta_R^{++}\,t^2(\theta_L^{++}\theta_R^{--} \theta_L^{--}\theta_R^{++}) = t^2
\ee
Thus, the $t$ dependent regularized pieces of the bosonic and fermionic integrations cancel, resulting in a constant.  This constant is determined from the supergravity matching of $Z_{BH}$ and $Z_{top}$ as 
in~\cite{0608021} and comes out as $g_{top}^{-2}$.
Using this in eqn. (\ref{zeroM}) and summing over genus $g$, the final result
is 
\be \label{ftop}
\sum_g g_{\rm top}^{2g-2} F_g, 
\ee
which is the full topological string free energy ${\mathcal F}_{\rm top}$, and $F_g$ are given in eqn. (\ref{topA}). $Z_{\rm top}$ corresponds
to the A-model topological string partition function, obtained from ${\mathcal F}_{\rm top}$
in eqn. (\ref{ztop1}). 

For the case of anti-instanton, one uses $u = 1/z$ and localizes the integration region to $|w|=0,|u|=0$, i.e., the center
of $AdS_2$ and the north pole of $S^2$. 
The final result using eqn. (\ref{zeroAM}), and summing over genus $g$ comes out as:
\be \label{barftop}
\sum_g g_{\rm top}^{2g-2}{\bar F}_{g}
\ee
which is the anti-topological string free energy ${\bar {\mathcal F}}_{\rm top}$ and $\bar F_g$ are given in eqn. (\ref{topAbar}). Its exponentiation, gives the
partition function ${\bar Z}_{\rm top}$. 

Therefore, the full type IIA partition function and hence, the black hole partition function is obtained from ${\mathcal F}_{\rm top}$ in (\ref{ftop}) and ${\bar {\mathcal F}}_{\rm top}$ in (\ref{barftop}), coming from instantons and anti-instantons respectively, to be $|Z_{\rm top}|^2$ and one recovers the relation (\ref{osv})
as $Z_{\rm BH} = Z_{IIA} = |Z_{\rm top}|^2$. We note that space-time supersymmetry was not used explicitly in the calculation of the type IIA partition function
and the subsequent emergence of topological string partition function. The localization procedure relied completely on world-sheet superconformal invariance and 
the BRST method. 
It is desirable to extend the computations of
this work towards ${\mathcal N}=4$ and ${\mathcal N}=8$ superstring theories as well. Applications towards other extensions of the OSV conjecture~\cite{Saraikin:2007jc}, are also important to 
explore.

\vskip 1.0cm

{\bf Acknowledgments}: I thank Nathan Berkovits for suggesting the problem, helpful discussions, for 
sharing his insights at crucial stages and for a careful reading of the draft. I thank the FAPESP grant 05/52888-9 for financial aid and IOP, Bhubaneswar for hospitality. 
I also thank Kuntala Bhattacharjee for her 
support during the course of this work.



\end{document}